\documentclass[sigconf,nonacm]{acmart}
\PassOptionsToPackage{table}{xcolor}
\AtBeginDocument{%
  }

\setcopyright{none}
\usepackage{xspace}
\usepackage{enumerate}

\usepackage{enumitem}
\usepackage{ocgx2}

\usepackage{xcolor}
\usepackage{listings}
\usepackage{tcolorbox}
\tcbuselibrary{listings,skins}

\definecolor{codeBlue}{HTML}{315A8A}
\definecolor{codeGreen}{HTML}{4F7D5C}
\definecolor{codeRed}{HTML}{9A4D55}
\definecolor{codeInk}{HTML}{243447}
\definecolor{codeMuted}{HTML}{8492A6}

\newcommand{\codeverb}[1]{\textcolor{codeBlue}{\texttt{\textbf{#1}}}}
\newcommand{\codetarget}[1]{\textcolor{codeBlue}{\texttt{\textbf{#1}}}}
\newcommand{\codeparam}[1]{\textcolor{codeGreen}{\texttt{#1}}}
\newcommand{\codevalue}[1]{\textcolor{codeRed}{\texttt{#1}}}

\lstdefinelanguage{codeJavaScript}{
  morekeywords={async,await,break,case,catch,class,const,continue,debugger,default,delete,do,else,export,extends,false,finally,for,function,if,import,in,instanceof,let,new,null,return,super,switch,this,throw,true,try,typeof,var,void,while,with,yield},
  morekeywords=[2]{chart,createLinePlot,editLineMark,createCanvas,createData,createScatterPlot,encodeColor,editPointMark,createRegression,createGuides,render,querySelector,getContext},
  sensitive=true,
  morecomment=[l]{//},
  morecomment=[s]{/*}{*/},
  morestring=[b]',
  morestring=[b]"
}

\lstdefinestyle{codeJavaScript}{
  language=codeJavaScript,
  basicstyle=\ttfamily\fontsize{8}{9.6}\linespread{1}\selectfont\setlength{\baselineskip}{1.2em}\color{codeInk},
  keywordstyle=\color{codeBlue}\bfseries,
  keywordstyle=[2]\color{codeBlue}\bfseries,
  identifierstyle=\color{codeGreen},
  commentstyle=\color{codeGreen}\itshape,
  stringstyle=\color{codeRed},
  numbers=left,
  numberstyle=\scriptsize\color{codeMuted},
  numbersep=4pt,
  xleftmargin=0pt,
  resetmargins=true,
  lineskip=-0.66pt,
  breaklines=true,
  breakatwhitespace=false,
  columns=fullflexible,
  keepspaces=true,
  showstringspaces=false,
  tabsize=4,
  captionpos=b,
  aboveskip=0.7\baselineskip,
  belowskip=0.5\baselineskip,
  literate=*
    {0}{{{\color{codeRed}0}}}1
    {1}{{{\color{codeRed}1}}}1
    {2}{{{\color{codeRed}2}}}1
    {3}{{{\color{codeRed}3}}}1
    {4}{{{\color{codeRed}4}}}1
    {5}{{{\color{codeRed}5}}}1
    {6}{{{\color{codeRed}6}}}1
    {7}{{{\color{codeRed}7}}}1
    {8}{{{\color{codeRed}8}}}1
    {9}{{{\color{codeRed}9}}}1
}

\newtcblisting{codeblock}{
  listing engine=listings,
  listing only,
  listing options={style=codeJavaScript},
  hbox,
  enhanced,
  frame hidden,
  interior hidden,
  boxsep=0pt,
  left=0pt,
  right=0pt,
  top=0pt,
  bottom=0pt,
  before={\begin{center}},
  after={\end{center}}
}

\usepackage{array,tabularx,colortbl,ragged2e,zref-savepos}
\newcolumntype{Z}[1]{>{\hsize=#1\hsize\linewidth=\hsize\RaggedRight\arraybackslash}X}

\def\lib{\textsf{ggaction}\xspace}

\newcommand{\myparagraph}[1]{\vspace{3pt}\noindent\textbf{#1.}}

\newcommand{\myparagraphit}[1]{\vspace{3pt}\noindent\textit{#1.}}

\begin{document}

\title{ggaction: A Grammar of Graphical Actions}

\author{Hyeon Jeon}
\email{hj@cs.au.dk}
% \correspondingauthor
\affiliation{%
  \institution{Seoul National University}
  \city{Seoul}
  % \state{}
  \country{Korea}
}

\author{Jinwook Seo}
\authornote{Corresponding author.}
\email{jseo@snu.ac.kr}
% \correspondingauthor
\affiliation{%
  \institution{Seoul National University}
  \city{Seoul}
  % \state{}
  \country{Korea}
}

% \affiliation{%

% \affiliation{%

% \affiliation{%

% \correspondingauthor
% \affiliation{%

\renewcommand{\shortauthors}{Jeon and Seo}

\begin{abstract}
 \textit{A chart may be declarative; authoring it is not.}
Visualization grammars often describe charts as finished specifications, whereas people construct them through a sequence of authoring actions. This mismatch can make visualization code difficult for humans to interpret and for machines to generate from human intent. 
\textit{\lib} addresses this gap by modeling the chart authoring process itself. 
In \lib, individual authoring actions are abstracted as functions, and the authoring process is expressed as a chain of these functions. 
This representation more closely aligns chart designers' authoring intent with code specifications, making the code easily understandable to both humans and machines, including language models.
Through a series of evaluations, we show that \lib is sufficiently expressive to capture common chart authoring intents and outperforms widely used visualization grammars, including Vega-Lite and ggplot2, in both human and machine interpretability.
\lib is available at \href{https://github.com/ggaction/ggaction}{github.com/ggaction/ggaction}.

% We evaluate \lib against widely used visualization grammars, including Vega-Lite and ggplot2, by comparing the degree to which machines and human progammers can understand 

% the complexity of authoring charts and building chart-authoring support systems using each grammar.

\end{abstract}




% Static final frame for compatibility across PDF viewers.
\begin{teaserfigure}
  \centering
  \includegraphics[width=\textwidth]{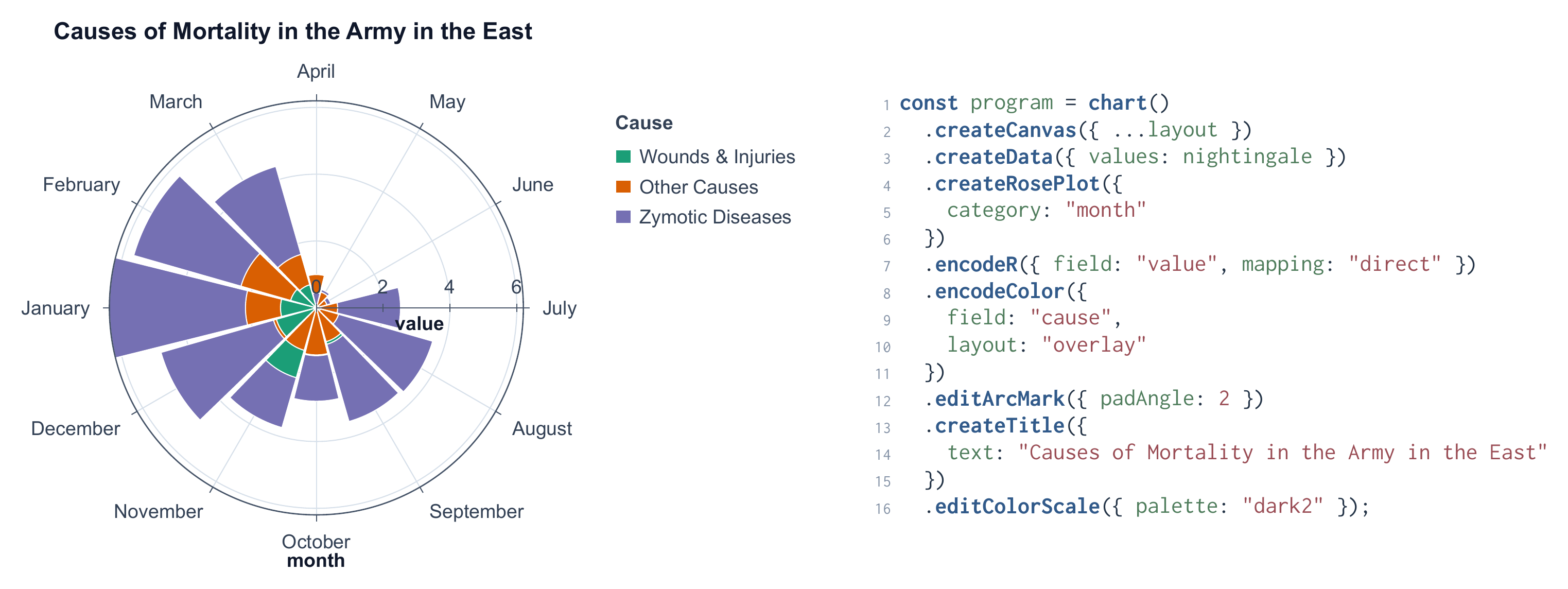}
  \caption{A rose plot (also known as a Nightingale plot) specified by \lib (left) and the corresponding code snippet (right). In \lib, a chart is defined as a sequence of authoring actions, allowing users (and language models) to easily translate authoring intent to code.}
  \label{fig:teaser}
\end{teaserfigure}

\received{20 February 2007}
\received[revised]{12 March 2009}
\received[accepted]{5 June 2009}

\maketitle

\section{Introduction}

Visualization authoring is inherently sequential and procedural \cite{jeon26tvcg, parsons22tvcg, shin25tvcg, Munzner2014, yunwang23tvcg}.
Designers typically begin with a vague objective, progressively detail and refine the chart through a sequence of design actions, and ultimately converge on an acceptable solution \cite{jeon26tvcg}.
Even constructing a simple monochrome scatterplot involves a series of authoring actions, including selecting a dataset and chart type, mapping data fields to the $x$- and $y$-axes, and specifying chart and mark sizes.

However, visualization grammars are typically declarative \cite{satyanarayan17tvcg, wickham2011book, heer10tvcg, mcnutt23tvcg}.
These grammars, such as Vega-Lite \cite{satyanarayan17tvcg} and ggplot2 \cite{wickham2011book}, represent charts as finished specifications that describe only the final visual outcome, hardly capturing the process by which a chart is constructed.
This mismatch between declarative grammars and the procedural nature of chart authoring \cite{satyanarayan20tvcg, jeon26tvcg, liu21vis, liu25tvcg} makes it difficult for both machines and humans to interpret specifications and design charts. 
From a machine perspective, this mismatch makes it harder for agents, typically built on language models, to translate users' chart authoring intent (often expressed as a sequence of natural language prompts, each corresponding to an individual authoring action) into code that specifies a visualization.
From a human perspective, designers may find declarative grammars difficult to interpret and to relate to the resulting charts \cite{liu21vis, hoffswell18chi}.
Procedural grammars have been proposed to address this limitation \cite{liu21vis, liu25tvcg}, but they largely focus on transforming graphical objects (i.e., changing mark sizes or color) rather than high-level chart semantics, leaving a gap between their abstractions and users' actual authoring intents.

% makes it difficult for both machines and humans to understand and reason about the authoring process. 

% thereby discarding information about how the chart is constructed. For example, even when a user repeatedly changes the color palette, this design history is not reflected in the final specification. 
% This mismatch between declarative grammars and the procedural nature of chart authoring \cite{satyanarayan20tvcg, jeon26tvcg, liu21vis} makes it difficult for both machines and humans to understand and reason about the authoring process. 
% From a machine perspective, this mismatch complicates the design and implementation of chart-authoring systems \cite{wu2020visact}, often requiring additional layers that translate authoring actions into declarative specifications \cite{hyeok22chi, wang23tvcg, jeon26tvcg}. From a human perspective, declarative grammars may be difficult for designers to interpret and relate to the charts they generate \cite{liu21vis, hoffswell18chi}.

\begin{figure*}[p]
    \centering
    \resizebox{\textwidth}{!}{%
        \rotatebox{-90}{%
            \begin{minipage}{\textheight}
                \centering
                \includegraphics[width=\linewidth]{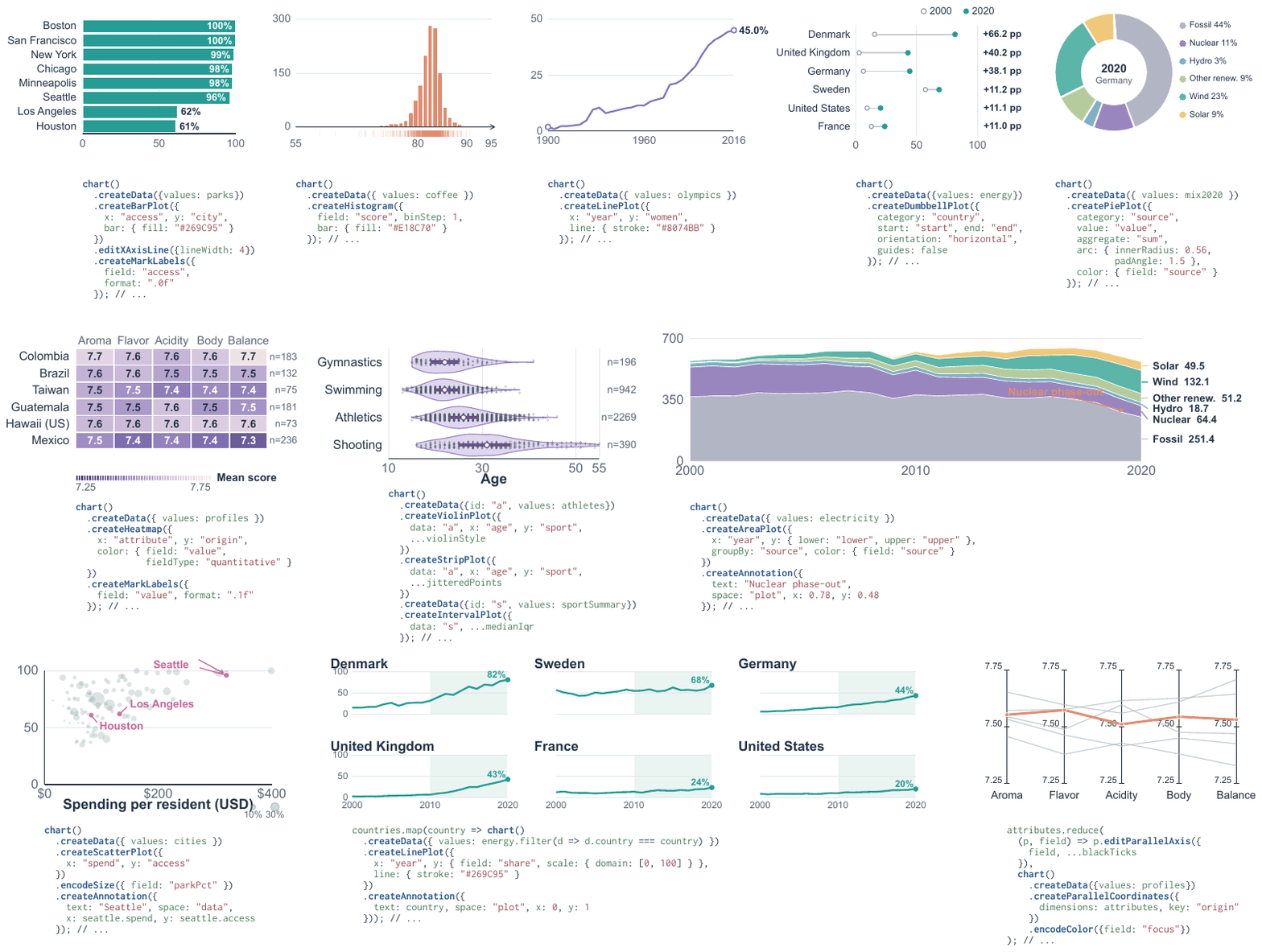}
                \caption{Example charts constructed using \lib and corresponding code specifications. In \lib, charts are specified as a sequence of authoring actions, making it easy to translate authoring intent into code.}
                \label{fig:gallery}
            \end{minipage}%
        }%
    }
\end{figure*}

To address this gap, we present \textit{\lib} (short for a grammar of graphical actions), a visualization grammar for specifying chart authoring processes. In \lib, each authoring action is represented as a function, and an authoring process is expressed as a chain of function calls. 
As users' authoring intent evolves, the corresponding specification can be extended by simply appending new function calls to this chain, rather than repeatedly revisiting and modifying code scattered throughout the specification (see \autoref{fig:gallery} for example charts and their corresponding code). 
This incremental structure supports effective and efficient visualization authoring for both humans and machines, such as language model-based programming agents. 
These actions span multiple levels of abstraction, from declaring simple charts (e.g., creating a histogram) to adjusting design details (e.g., changing the axis color). High-level actions are compiled into combinations of low-level actions. 
By supporting actions at multiple levels of granularity, \lib  preserves the expressiveness needed for specifying detailed chart design while enabling easy and efficient prototyping of simple charts.

We evaluate the utility of \lib from three perspectives.
First, we assess the expressiveness of \lib by analyzing its coverage of the taxonomy of chart authoring intent \cite{wang23tvcg}. 
Second, we measure how accurately language models interpret and generate \lib specifications, comparing it with alternative grammars, including Vega-Lite and ggplot2. 
Finally, through a user study, we show that office workers with prior programming experience but relatively little experience with visualization grammars can understand \lib specifications with a smaller burden compared to Vega-Lite and ggplot2.
We conclude by discussing the future role of visualization grammars as code generation becomes increasingly automated and inexpensive.

\section{Background and Related Work}

Our work builds on three areas of prior work: understanding of the chart authoring process, visualization grammars, and authoring with agents.

\subsection{Chart Authoring Process}

\label{sec:chartauthoringprocess}

We discuss the different perspectives that view the chart authoring process.

\myparagraph{Authoring as an automated design process}
Early work on chart authoring assumes users have a clear understanding of their objectives or design intent, where these intents can serve as input to automated recommendation systems \cite{mackinlay86tog, mackinlay07tvcg, roth90chi, moritz19tvcg, lin20chi, shen21eurovis}.
% Early work on visualization authoring focuses on automating the search for effective charts for current data and requirements \cite{mackinlay86tog, mackinlay07tvcg, roth90chi, moritz19tvcg, lin20chi, shen21eurovis}. 
% Here, chart designers are assumed to have a clear understanding of their objectives or design intent, which can then serve as input to automated recommendation systems.
At the outset, Mackinlay's APT \cite{mackinlay86tog} formalizes visualizations as compositions of primitive operators (e.g., position, color, and axes) and uses this formalism to search for effective visualizations given a particular data context. 
This approach has been extended in various directions over decades; for example, while Roth and Mattis \cite{roth90chi} attempt to characterize input data context in more detail to inform automated systems, Draco \cite{moritz19tvcg} aims to formalize design knowledge itself as constraints for finding effective visualizations.

% This perspective has also influenced diverse production-level chart authoring systems. For example, Show Me \cite{mackinlay07tvcg} was integrated into Tableau and has been widely adopted in practice. Widely used commercial tools like Microsoft Excel or Spotfire also similarly incorporate automated recommendation algorithms to assist users in chart authoring.

\myparagraph{Authoring as an iterative design process}
In parallel, the literature has recognized chart authoring as an iterative process \cite{bigelow14avi, satyanarayan20tvcg}.
This perspective models chart design as a series of incremental construction, inspection, and revision of visual representations to reach the designer's intent \cite{bigelow14avi, stolte02tvcg, mckenna14tvcg}.
Reflecting this view, authoring systems such as Lyra \cite{satyanarayan14cgf, zong21tvcg} and Charticulator \cite{ren19tvcg} enable users to progressively develop visualization designs through direct manipulation. Meanwhile, Bigelow et al. \cite{bigelow17tvcg} facilitate iterative authoring across multiple tools by enabling designers to bridge otherwise disconnected authoring tools (e.g., D3 \cite{bostock11tvcg} and Illustrator).
Shin et al. \cite{shin23perceptualpat, shin25visualizationary} demonstrate that feedback grounded in the iterative authoring process can support designers in designing effective charts.

\myparagraph{Authoring as a provisional design process}
Recent studies suggest that chart design is not merely iterative, but also situated and provisional \cite{parsons22tvcg, alspaugh19tvcg, parsons16tvcg, parsons26arxiv}. Designers may begin without a clear understanding of what they should create---or even the problem they are trying to address---and progressively articulate both through trial and error \cite{alspaugh19tvcg, parsons26arxiv}. 
For instance, Parsons et al. \cite{parsons20vis, parsons26arxiv2} show that design does not proceed according to a fixed set of rules; rather, designers exercise judgment by interpreting the current situation and drawing on prior experiences with similar situations. They also show that designers may not even know what to do next until they try; instead, they probe different possibilities in advance and interpret the outcomes to determine how the design should proceed \cite{parsons26arxiv}.

\myparagraphit{Our contribution}
Our understanding of chart authoring has evolved from an automated design problem to an iterative and, more recently, situated and provisional process. However, many visualization grammars specify only the final visualization, reflecting an earlier, outcome-centric view of authoring. 
\lib addresses this gap by representing a chart as a specification of the authoring actions.

\subsection{Visualization Grammars}

We categorize and examine existing visualization grammars along two dimensions: declarative and procedural.

\myparagraph{Declarative grammars}
Declarative grammars describe charts as compositions of graphical components. 
They typically define the high-level semantic structure of a visualization through the relationships of such components, specifying \textit{what} a chart consists of. 
A seminal example is Wilkinson's Grammar of Graphics (GoG) \cite{wilkinson12gog}, which characterizes charts using seven classes of components, including data transformations, scales, geometric objects, and visual aesthetics. 
A key strength of this approach is its expressiveness: rather than defining each chart type as a separate primitive, a grammar can generate a broad range of visualizations by composing a relatively small set of reusable components.

Building on this principle, many high-level visualization grammars adopt declarative representations. Wickham \cite{wickham2011book}, for example, extends the GoG into a layered grammar of graphics, which underlies the ggplot2 grammar \cite{Wickham2016ggplot}. 
Declarative abstractions are subsequently developed for web-based or interactive visualization, including Protovis \cite{heer10tvcg, bostock09tvcg}, Plotly \cite{plotly}, Vega, and Vega-Lite \cite{satyanarayan17tvcg}. 
 % allows authors to write low-level chart specifications and serves as a backend for Vega-Lite. 
More specialized declarative grammars have also been proposed \cite{mcnutt23tvcg}, e.g., for multiclass scatterplots \cite{jo19tvcg}, genomic data \cite{lyi22tvcg}, and data sonification \cite{kim24chi}.

\myparagraph{Procedural grammars}
In contrast, procedural chart grammars specify charts through a sequence of operations that create and modify them.
Matplotlib \cite{hunter07cse} and Prefuse \cite{heer05chi}, for example, support stepwise chart construction and customization through a series of plotting and editing operations.
D3 \cite{bostock11tvcg} provides even finer-grained control, allowing authors to select data and graphical objects, bind data to them, and directly modify their visual properties.
These systems give authors control over low-level visual details but often require them to reason about the underlying graphical objects and execution mechanisms (e.g., DOM in web technologies).

Later work introduced higher-level procedural abstractions. Atlas \cite{liu21vis}, for example, constructs visualizations by repeating or dividing graphical elements, enabling complex charts to be built through sequences of transformations. 
Manipulable Semantic Components \cite{liu25tvcg} further raises the level of abstraction by representing charts as visualization-specific objects, such as marks, collections, and encodings, along with operations to create and modify them.
Collectively, these endeavors shift procedural visualization languages from low-level graphical manipulation toward operations that capture semantically meaningful visualization and authoring decisions.

\myparagraphit{Our contribution}
\lib leverages procedural abstractions, but organizes its grammar explicitly around \textit{authoring actions}: decisions that users may naturally make while constructing and refining a chart in the real world.
This principle contrasts with previous procedural grammars that treat operations as transformations of graphical or chart semantics, such as marks and channels.
These actions are organized hierarchically, such that higher-level actions (e.g., creating a scatterplot) are composed of lower-level ones (e.g., editing axis ticks), allowing users to express multiple levels of authoring intent---which is common in chart authoring in practice \cite{jeon26tvcg}.
Moreover, \lib represents a chart program as a linear sequence of progressively accumulated actions, allowing subsequent design decisions to be expressed as localized additions to the existing program. This design improves the interpretability of chart authoring for both humans and machines, facilitating more efficient authoring and the development of authoring support tools.

% Together, these design choices make authoring intent an explicit organizing principle of the visualization grammar.

% This compositional model provides a general framework for specifying both conventional chart types and novel variations within the same language.

\subsection{Chart Authoring with Agents}

We design \lib in the context of language model-based assistants that interpret authoring intent and translate it into executable code. We review prior work on chart authoring through the lens of how this translation is executed.

\myparagraph{Directly synthesizing programs}
A major branch of agent-supported chart authoring directly synthesizes chart specifications from natural language intents.
Early systems such as Chat2Vis \cite{maddigan11access} and LIDA \cite{dibia23acl} demonstrate this approach by prompting language models with natural language requests and data context to generate visualization code. 
Subsequent work has made this process more robust by leveraging iterative debugging and feedback \cite{yang24aclfindings, koh25naacl, goswami25www, chen26iclr, lu26sigmod}.

% . For example, Koh et al. \cite{} propose to provide feedback on automatically generated charts, while more recent work has incorporated multimodal reasoning and multi-agent workflows to further refine the generation process \cite{}.

Another notable direction is to train small models specifically for visualization generation. Although frontier large language models (LLMs) can generate visualization code effectively, relying on them may be impractical due to limited access and high computational and environmental cost. 
Recent work thus fine-tunes small language models (SLMs) for chart generation \cite{voigt24coling, ni25acl}. For example, Voigt et al. \cite{voigt24coling} train SLMs to generate Vega-Lite specifications from data and natural language prompts, whereas Pesaran et al. \cite{pesaran24emnlp} specialize models for generating Matplotlib code.

\myparagraph{Progressively transforming specifications}
Whereas the works discussed above largely translate chart authoring intent into a complete code specification in a monolithic manner, another branch adopts a more fine-grained approach, progressively incorporating users' authoring intents to evolve the visualization specification over time. Many recent authoring systems follow this paradigm.
For example, DynaVis \cite{Vaithilingam24chi} and TailVis \cite{song26arxiv} allow users to iteratively modify charts via natural language input while exposing widgets for subsequent refinement, enabling later changes through lightweight parameter adjustments. L'Yi et al. \cite{lyi25tvcg} similarly propagates generated modifications to other authoring interfaces, helping users understand the changes and continue editing through alternative interaction modalities. DirectVis \cite{park26pacificvis}, meanwhile, captures users' authoring intents through direct manipulation and translates those interactions into corresponding code modifications. Visualization Autocomplete \cite{jeon26tvcg} predicts potential authoring intents before users fully specify them, thereby reducing the effort required to progressively construct charts.

\myparagraph{Leveraging intermediate representations}
The difficulty of directly translating chart authoring intents into visualization code has also motivated approaches that introduce intermediate representations for expressing intents before generating executable specifications.
Setlur et al. \cite{setlur19iui} translate natural language intents into an intermediate representation called ArkLang, which is then used to issue queries to a visual analytics system.
Similarly, Narechania et al. \cite{narechania21tvcg} represent users' analytical intents as an intermediate specification and use it to generate Vega-Lite code.

Another branch explores generating such intermediate specifications with machine learning models, including language models \cite{ouyang25acl, luo22tvcg, wu24sigmod}.
For example, Luo et al. \cite{luo22tvcg} introduce Vega-Zero, a linearized representation of Vega-Lite designed to make specifications easier for sequence models to generate.
Wang et al. \cite{wang23tvcg} represent chart authoring intents as combinations of operations, objects, and parameters, which are then executed to modify the charts.
More recently, Flint \cite{wang26arxiv} and Raiven \cite{irger26arxiv} further raise the level of abstraction: they let the model express authoring intent through a higher-level grammar and delegate the translation into executable specifications to a compiler.

\myparagraphit{Our contribution}
Prior work has explored diverse approaches for translating authoring intent into chart specifications.
However, this translation is often monolithic, and
even systems that support progressive specification updates do not provide a reusable underlying abstraction for such transformations. Prior work has also noted that the lack of such abstractions makes these systems difficult to develop and extend \cite{jeon26tvcg}.
\lib addresses this gap by making authoring actions the basic units of chart programs.
We investigate whether this alignment between authoring intent and program structure improves both human and machine interpretability.

% Collectively, prior work improves intent-to-code translation by strengthening program synthesis, transforming existing specifications, or introducing intermediate representations.
% However, these approaches still leave a representational gap between designers' localized authoring intents and the structure of executable chart programs: models must either reason in the target grammar, translate each intent into specification-level edits, or rely on an additional representation that is subsequently mapped to an underlying visualization system.
% \lib addresses this gap by making authoring actions the basic units of the executable chart program itself.
% These actions compose hierarchically and accumulate progressively, allowing successive authoring intents to be expressed as localized additions while delegating their lower-level implementation to the grammar.
% We investigate whether this alignment improves both machine and human interpretability.

% \lib complements these approaches by organizing executable chart programs around authoring actions.
% These actions compose hierarchically and accumulate sequentially, allowing successive authoring intents to be expressed as localized additions to an existing program.
% We investigate how this alignment between authoring intent and program structure supports both machine-driven chart revision and human understanding.

\section{A Grammar of Graphical Actions}

We present \lib, a grammar that specifies charts as a sequence of authoring actions. 
% We first recap the motivation of \lib, then explain the grammar model of \lib. 
% Finally, we describe design features that support flexible, iterative chart authoring.

\subsection{Motivation}

Our core motivation is that existing visualization grammars do not align well with how users express authoring intent. 
Existing visualization grammars, such as Vega-Lite, Plotly, and seaborn, are primarily designed to describe the visual structure and representation of a chart. 
They are therefore well suited for specifying a completed visualization.

In practice, however, chart authoring is inherently iterative and progressive \cite{shin23perceptualpat, jeon26tvcg, parsons26arxiv2}. Rather than constructing a complete specification from the outset, users incrementally refine a chart through localized authoring intents, such as ``make the bars wider,'' ``sort by this column,'' or ``highlight this point.'' 
In existing grammars, such simple requests may translate into low-level code edits that span multiple, distant parts of the specification \cite{jeon26tvcg}. This mismatch makes the authoring process difficult to interpret not only for humans, but also for LLMs, which process text in sequential streams and must therefore infer dependencies across distant parts of the code.

The misalignment also hinders the design and development of chart authoring support systems (e.g., \cite{Vaithilingam24chi, jeon26tvcg, ren19tvcg, song26arxiv}). 
These systems typically treat individual authoring actions as the basic unit of chart authoring, for example, by recommending appropriate actions \cite{jeon26tvcg} or providing interfaces tailored to specific actions \cite{Vaithilingam24chi, song26arxiv}.
Therefore, existing visualization grammars are difficult to adopt seamlessly in such systems. For example, Jeon et al. \cite{jeon26tvcg} note that applying a new authoring action to Vega-Lite is nontrivial: it often requires not only adding a new leaf node, but also coordinating changes across multiple parts of the specification. To bridge this mismatch, they implement an extensive set of functions that translate authoring actions into corresponding Vega-Lite patches.

Our new grammar, \lib, is motivated by this problem. 
The core design principle is to make chart edits explicit: when a user intends to modify a chart, the corresponding edit should be selectable from a predefined function pool. 
We also enable authoring actions to accumulate progressively, with later actions overriding earlier ones when necessary. This enables chart authoring systems to execute extended authoring sequences without requiring additional engineering efforts. In the following section, we describe the design of \lib in detail.

\subsection{Grammar Model}

\label{sec:grammarmodel}

We discuss the design model of \lib. 
To represent users' authoring intent as simply as possible, \lib adopts only two levels of abstraction. At the lower level, \textit{graphical actions} capture individual chart editing intents. At the higher level, a \textit{chart program} composes these actions as a complete chart. Note that we develop \lib as a JavaScript library to ease its integration with web-based environments.

\myparagraph{Graphical actions}
The basic unit of chart specification in \lib is the authoring action. This design reflects the observation that users often express their authoring intent through localized actions \cite{parsons22tvcg, parsons26arxiv2, jeon26tvcg, srinivasan21chi}. 
Thus, by adopting actions as its basic unit, \lib more closely aligns the code semantics with the way users formulate authoring intent.

We abstract each authoring action to an individual function, following the notion of Wang et al. \cite{wang23tvcg}.
In detail, each action is named using a ``verb-noun'' pair. The verb specifies the editing or design operation to be performed, while the noun identifies the chart element to which the operation is applied.
In addition, specific design decisions for that action are specified as function parameters.
For example, the intent \textit{``make the lines thicker''} can be expressed as:
\begin{codeblock}
editLineMark({strokeWidth: 4})
\end{codeblock}
where \codeverb{edit} specifies the operation, \codetarget{LineMark} identifies its target, and \codeparam{strokeWidth} specifies the parameters.

This design improves interpretability of \lib in three ways. 
First, it allows both human designers and language models to directly map design intent to code. Second, because all actions follow a consistent structure, even an unfamiliar action can be understood approximately from its name. 
Third, it provides strong locality: each authoring intent is encapsulated in a single function call.

\myparagraph{Chart program}
In \lib, each chart is represented as a \textit{program}, defined as a sequence of function calls, each corresponding to an authoring action.
The example chart program is as follows:
\begin{codeblock}
const program = chart()
  .createCanvas()
  .createData({ values: cars })
  .createScatterPlot({ x: "Displacement", y: "Acceleration" })
  .encodeColor({ field: "Origin" })
  .editPointMark({ opacity: 0.35 })
  .createRegression()
  .createGuides();
\end{codeblock}
This \codeparam{program} \codeverb{create}s a \codetarget{Scatterplot}, where the \codetarget{Color} of the \codetarget{Point Mark}s are \codeverb{encode}d based on the \codevalue{Origin} \codeparam{field}, and linear \codetarget{Regression} lines are drawn for each colored group.

The core design rationale here is to express chart programs as a linear, flat sequence of authoring actions. 
This contrasts with the nested structures used by many declarative grammars and reflects the sequential expression of intent in natural language. 
Although natural language has hierarchical syntactic structure, readers encounter intentions in sequence without explicitly reconstructing that hierarchy. 
By aligning chart programs with this sequential presentation, \lib aims to make code easier to follow as a progression of design decisions. The same alignment may also help language models translate natural language instructions, provided in turn, into corresponding sequences of authoring actions. 
This representation also makes the progression of design decisions explicit. In the code example above, readers can follow how the program constructs a scatterplot, distinguishes groups by color, reduces point opacity, and adds regression lines. 
A chart program thus communicates both what the chart contains and how its design decisions iteratively build on one another. 

After defining a chart program, users can render it by executing:
\begin{codeblock}
const canvas = document.querySelector("#chart");
render(program, canvas.getContext("2d"));
\end{codeblock}

Note that this example assumes a situation in which users want to render a chart in the web browser using canvas, while \lib also supports exporting the chart as a separate file (e.g., .svg or .pdf).
The rendered results will be:

\begin{figure}[h]
  \centering
  % \vspace{-7mm}
  \includegraphics[width=0.8\linewidth]{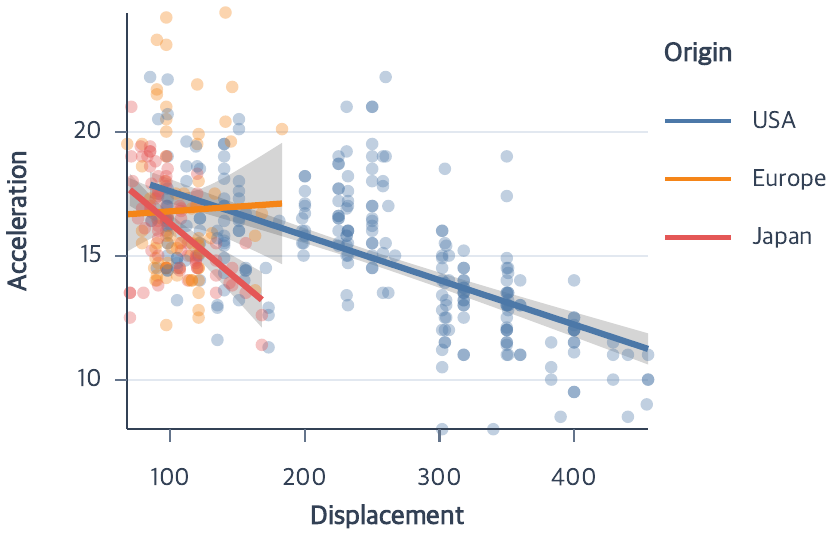}
  % \vspace{-7mm}
\end{figure}

\noindent

\subsection{Designing Authoring Actions}

\label{sec:designaction}

When designing \lib, selecting which authoring actions to include in the library is an important decision. We followed two principles:
\begin{itemize}[leftmargin=9pt]
\item \textit{Meaningfulness:} An action should correspond to a design decision that users may meaningfully make on its own.
\item \textit{Atomicity:} An action should capture a single design decision, rather than bundling multiple independently meaningful decisions.
\end{itemize}
Together, these principles bound the abstraction level of the action portfolio. 
Meaningfulness prevents actions from becoming unnecessarily fine-grained, which would increase the size and complexity of the vocabulary without reflecting how users formulate authoring intent. 
Conversely, atomicity prevents actions from becoming overly coarse-grained by ensuring that independent design decisions remain separate. 
An action can still coordinate complex changes across multiple chart components, but those changes should collectively realize a single authoring decision.

\begin{figure*}
    \centering
    \includegraphics[width=\linewidth]{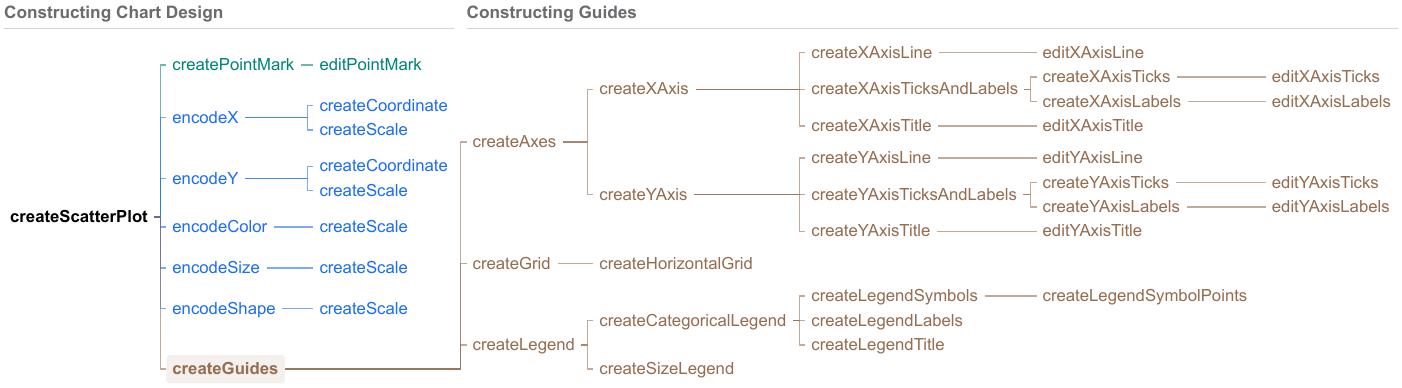}
    \caption{The hierarchical composition of \codeverb{createScatterPlot} in \lib. When invoked, this action internally decomposes into lower-level actions that specify the mark, encodings, and guides.}
    \label{fig:hier}
\end{figure*}

Within this bound, we design \lib to support authoring intents at multiple levels of abstraction (\autoref{fig:hier}). 
At the highest level, \lib provides actions like selecting chart types (e.g., \codeverb{createBarPlot} and \codeverb{createScatterPlot}). At the lowest level, it provides fine-grained actions for modifying individual visual properties (e.g., \codeverb{editXAxisTicks} with \codeparam{length} specified). 
We link these actions hierarchically: a higher-level action is defined as a sequence of lower-level actions.

This design allows \lib to preserve the simplicity of chart programs without sacrificing expressiveness. 
Users can invoke fine-grained actions only when specific revisions are needed, while relying on higher-level actions to establish the overall chart structure and avoid unnecessary function calls. 
This reflects the common authoring flow of visualization designers in practice \cite{jeon26tvcg, yunwang23tvcg}.
For example, when users want to create a scatterplot and modify only its opacity encoding, they need not explicitly specify the remaining encodings (e.g., through \codeverb{encodeX} or \codeverb{encodeY}). Instead, they can first call \codeverb{createScatterPlot} to establish the basic chart structure and then chain \codeverb{encodeOpacity} to refine the desired visual property.
This design also makes the length of a chart specification grow roughly in proportion to the amount of authoring intent expressed, avoiding unnecessary code expansion and thus improving interpretability.

This structure also helps both human designers and language models modify chart programs with fewer unintended errors. 
When a simple revision requires modifying or restating unrelated parts of the specification, it may increase the risk of inadvertently changing unaffected design decisions. In \lib, authors and language models can both express only the intended modification by appending or revising the corresponding fine-grained action, while leaving unrelated decisions untouched.

\subsection{Supporting Progressive Authoring Sequences}

Chart authoring is inherently iterative and even provisional: new design intent often emerges after an initial chart has been specified (\autoref{sec:chartauthoringprocess}).
Although users can express such intents by modifying an earlier action, doing so obscures the distinction between the original decision and its later revision.
\lib instead encourages users to append a new action to the existing chart program.
Each appended action operates on the state produced by preceding actions and returns a new immutable program, preserving the order in which design decisions were introduced.

For example, the following program creates a line chart:
\begin{codeblock}
const program = chart()
  .createCanvas()
  .createData({ values: cars })
  .createLinePlot({
    x: { field: "Year", fieldType: "temporal" },
    y: { field: "Acceleration", aggregate: "mean" },
  });
\end{codeblock}
Suppose users later decide to make the line thicker.
One way is to add the parameter \codeparam{line: \{ strokeWidth: }\codevalue{4 }\codeparam{\}} by revising the call to \codeverb{createLinePlot}.
However, instead, \lib recommends appending an \codeverb{editLineMark} action:
\begin{codeblock}
const revised = program.editLineMark({
  strokeWidth: 4
});
\end{codeblock}
Both approaches produce the same visual result, but the latter preserves the revision as an explicit step in the authoring sequence.
The resulting program therefore captures not only the current chart specification but also how it evolved.

Moreover, to support this progressive authoring pattern, \lib resolves repeated edits through ordered precedence: when multiple actions assign the same property, the latest action takes precedence.
Consider the following sequence:
\begin{codeblock}
const program = chart()
  ... // action sequence
  .editPointMark({ opacity: 0.35 })
  ... // action sequence
  .editPointMark({ opacity: 0.7 });
\end{codeblock}
Here, the second \codeverb{editPointMark} action overrides the opacity value assigned by the first, setting an \codeparam{opacity} of \codevalue{0.7}.

This design benefits human readers, who can follow actions in their execution order, but is particularly valuable for machine authoring.
For example, in a conversational chart authoring system \cite{luo22tvcg, martins25access}, the system can generate only one or a few actions that capture the newly expressed intent at each turn, rather than regenerating the entire specification. This property can simplify the design and implementation of such authoring systems. Moreover, because the chart is represented as a sequence of actions, the system can naturally return to intermediate states, undo specific actions, or explore alternative branches---capabilities that are increasingly important in modern chart authoring systems \cite{jeon26tvcg, song26arxiv, ueno26arxiv}.
\section{Technical Details}

\label{sec:techdetail}

We discuss the technical details involved in executing and developing \lib.

\subsection{Chart States}

A \lib action is invoked on a chart program and returns a new program whose state is updated according to the action.
We focus on three components of the program state: \texttt{semanticSpec}, \texttt{graphicSpec}, and \texttt{context}.
The \texttt{semanticSpec} represents the semantic structure of the chart, including datasets, layers, encodings, scales, coordinates, guides, and titles.
For actions that alter this semantic structure, \lib first updates the corresponding semantic state.
Then, the action materializes the resulting changes into \texttt{graphicSpec}, which represents the backend-agnostic graphical state consumable by a renderer.
It contains graphical objects such as rectangles, circles, lines, and text, together with their resolved geometry and appearance---for example, where an object is positioned and how it is colored.
Finally, \texttt{context} maintains information about the current authoring focus, such as the currently selected dataset, mark, or scale.
This state allows subsequent actions to infer omitted targets when they are unambiguous, reducing the burden on users of providing such information through boilerplate.
For example, when users add multiple datasets to a single \lib program, the most recently added dataset is treated as the currently active dataset by \texttt{context}.

We decouple \texttt{semanticSpec} from \texttt{graphicSpec} to simplify rendering. Renderers operate only on \texttt{graphicSpec}, without requiring access to chart semantics. Thus, \lib can more easily support multiple rendering backends, such as SVG, PNG, and PDF.
At the same time, these two specifications remain complementary. 
Maintaining both allows \lib to use semantic relationships to identify which graphical elements are affected by an action and selectively rematerialize only those elements. Storing only graphical state may make semantic edits difficult to resolve; for example, after a request such as ``change the x scale,'' it would be difficult to recover which marks, axes, and grid lines depend on that scale. 
Conversely, storing only semantic state may require translating the entire specification into graphical elements after every action, reducing the efficiency of the systems relying on \lib.

Updates to \texttt{semanticSpec} and \texttt{graphicSpec} are implemented through three lowest-level actions: \codeverb{editSemantic}, \codeverb{createGraphics}, and \codeverb{editGraphics}. \codeverb{editSemantic} modifies chart semantics, while \codeverb{createGraphics} and \codeverb{editGraphics} create and modify graphical objects, respectively.
As described earlier (\autoref{sec:designaction}), every action in \lib decomposes into lower-level actions. Recursively applying this decomposition eventually reduces any \lib action to a sequence of these three primitives. This design provides \lib with a small, uniform execution core.
It is worth noting that, although users may invoke these primitive actions directly, \lib aims to have most charts authored through combinations of higher-level actions, without requiring users to manipulate these low-level operations.

\subsection{Rematerialization}

Because \texttt{graphicSpec} stores materialized graphical objects, a later action may invalidate graphics produced by earlier actions. For example, changing the domain of an $x$ scale affects not only the semantic state of the scale, but also the positions of marks and the corresponding axis and grid. We refer to the process of recomputing such affected graphical state as \textit{rematerialization}.

Rather than rematerializing the entire chart after every action, \lib identifies the graphical components that depend on the revised semantic state and selectively recomputes only those components. Each action is therefore responsible for propagating its changes to the relevant downstream graphical state.
This design allows a user-facing action to remain atomic while encapsulating the coordinated updates required across multiple graphical components, enabling users to focus on their higher-level intent without reasoning about downstream dependencies.

\subsection{LLM-Assisted Action Development}

As described above, implementing an action in \lib requires not only updating chart state, but also coordinating the rematerialization of graphical components affected by that update. Encapsulating this responsibility within each action simplifies chart authoring for users but makes it harder for developers to create new actions.

To manage this tradeoff, we adopted LLM-based programming as the primary workflow for developing new actions. 
Developers specify the expected behavior and API of an action, while coding agents implement the corresponding state management and rematerialization logic. 
To support this process, the \lib repository includes detailed agent instructions that define the responsibilities and implementation rules of each action.
Agents consult these instructions together with the action specification to produce implementations consistent with the existing codebase, reducing the need for developers to repeatedly restate common design principles and requirements.
Agents also generate tests that cover the rematerialization behavior when followed by other actions in diverse authoring sequences. 

However, decisions that affect the user-facing API or chart semantics remain subject to developer judgment and approval. The goal of this workflow is therefore not to replace human design decisions, but to allow developers to focus on what an action should express while delegating repetitive implementation. Notably, \lib itself was developed extensively through this workflow, using multiple generations of coding models, including OpenAI GPT-5.5, GPT-5.6 Sol, and GPT-6 Astra, using the Codex application.
\section{Evaluating Expressiveness}

\label{sec:expresiveness}

We evaluate the expressiveness of \lib by examining the degree to which the library covers the authoring intents.

\begin{table*}[tp]
  \captionsetup{skip=3pt}
  \caption{\lib's coverage of the 23 object categories in the taxonomy of Wang et al. \cite{wang23tvcg}. 21 objects are directly supported and two are partially supported.}
  \label{tab:object-coverage-1}
  \centering
  \renewcommand{\tabularxcolumn}[1]{m{#1}}
  \fontsize{5.6}{6.272}\selectfont
  \linespread{1}\selectfont
  \setlength{\tabcolsep}{1pt}
  \renewcommand{\arraystretch}{1.0}
  \setlength{\aboverulesep}{0.15ex}
  \setlength{\belowrulesep}{0.25ex}
  \rowcolors{2}{black!5}{white}
  \begin{tabularx}{\textwidth}{@{}Z{0.38} Z{0.80} l Z{1.61} Z{1.21}@{}}
    \toprule
    Object & Definition & Coverage & Direct & Partial \\
    \midrule
    Data field & Named data attribute. & Direct & \mbox{\texttt{createData}}, \mbox{\texttt{createTimeUnitData}}, \mbox{\texttt{createWindowData}}, \mbox{\texttt{encodeX}}, \mbox{\texttt{encodeY}} &  \\
    Data point & Individual records or selected members. & Direct & \mbox{\texttt{selectMarks}}, \mbox{\texttt{filterMarks}}, \mbox{\texttt{highlightMarks}} &  \\
    Data type & Declared or inferred field type. & Direct & \mbox{\texttt{encodeX}}, \mbox{\texttt{encodeY}}, \mbox{\texttt{encodeColor}}, \mbox{\texttt{encodeShape}}, \mbox{\texttt{encodeOpacity}} &  \\
    Data range & Contiguous or predicate-defined values. & Direct & \mbox{\texttt{filterData}}, \mbox{\texttt{filterMarks}}, \mbox{\texttt{encodeXRange}}, \mbox{\texttt{encodeYRange}} &  \\
    Chart & The complete visualization. & Direct & \mbox{\texttt{createScatterPlot}}, \mbox{\texttt{createLinePlot}}, \mbox{\texttt{createBarPlot}}, \mbox{\texttt{createHistogram}}, \mbox{\texttt{createHeatmap}}, \mbox{\texttt{createParallelCoordinates}} &  \\
    Canvas & The drawing or plot area. & Direct & \mbox{\texttt{createCanvas}}, \mbox{\texttt{editCanvas}} &  \\
    Mark & Data-encoding symbol or symbol set. & Direct & \mbox{\texttt{createPointMark}}, \mbox{\texttt{createLineMark}}, \mbox{\texttt{createBarMark}}, \mbox{\texttt{createAreaMark}}, \mbox{\texttt{createRectMark}}, \mbox{\texttt{createRuleMark}} &  \\
    Axis & Axis, including labels and ticks. & Direct & \mbox{\texttt{createAxes}}, \mbox{\texttt{createXAxis}}, \mbox{\texttt{createYAxis}}, \mbox{\texttt{editXAxis}}, \mbox{\texttt{editYAxis}}, \mbox{\texttt{editRadialAxis}} &  \\
    Title & Chart title or title text. & Direct & \mbox{\texttt{createTitle}}, \mbox{\texttt{editTitle}}, \mbox{\texttt{removeTitle}} &  \\
    Legend & Chart legend or key. & Direct & \mbox{\texttt{createLegend}}, \mbox{\texttt{editLegend}}, \mbox{\texttt{editLegendLayout}}, \mbox{\texttt{editLegendLabels}}, \mbox{\texttt{editLegendTitle}}, \mbox{\texttt{removeLegend}} &  \\
    Gridline & Coordinate-system grid line. & Direct & \mbox{\texttt{createGrid}}, \mbox{\texttt{editGrid}}, \mbox{\texttt{removeGrid}}, \mbox{\texttt{createHorizontalGrid}}, \mbox{\texttt{createVerticalGrid}}, \mbox{\texttt{createThetaGrid}} &  \\
    Mark channel & Data-bindable visual channel. & Direct & \mbox{\texttt{encodeX}}, \mbox{\texttt{encodeY}}, \mbox{\texttt{encodeColor}}, \mbox{\texttt{encodeGroup}}, \mbox{\texttt{encodeSize}}, \mbox{\texttt{encodeShape}} &  \\
    Opacity & Element transparency. & Direct & \mbox{\texttt{encodeOpacity}}, \mbox{\texttt{editPointMark}}, \mbox{\texttt{editLineMark}}, \mbox{\texttt{editBarMark}}, \mbox{\texttt{editAreaMark}}, \mbox{\texttt{editTextMark}} &  \\
    Stroke & Line or border appearance. & Direct & \mbox{\texttt{encodeStroke}}, \mbox{\texttt{encodeStrokeWidth}}, \mbox{\texttt{encodeStrokeDash}}, \mbox{\texttt{editLineMark}}, \mbox{\texttt{editGrid}}, \mbox{\texttt{editLegendBorder}} &  \\
    Text font & Text font properties. & Direct & \mbox{\texttt{editTextMark}}, \mbox{\texttt{editTitle}}, \mbox{\texttt{editLegendLabels}}, \mbox{\texttt{editLegendTitle}}, \mbox{\texttt{editXAxisLabels}}, \mbox{\texttt{editYAxisLabels}} &  \\
    Text color & Text color. & Direct & \mbox{\texttt{editTextMark}}, \mbox{\texttt{editTitle}}, \mbox{\texttt{editLegendLabels}}, \mbox{\texttt{editLegendTitle}}, \mbox{\texttt{editXAxisLabels}}, \mbox{\texttt{editYAxisLabels}} &  \\
    Label & Data or series label. & Direct & \mbox{\texttt{createMarkLabels}} &  \\
    Annotation text & Free-form explanatory text. & Direct & \mbox{\texttt{createAnnotation}} &  \\
    Reference line & Reference-value line; averages counted separately. & Direct & \mbox{\texttt{createReferenceLine}} &  \\
    Reference band & Band spanning a reference range. & Direct & \mbox{\texttt{createReferenceBand}} &  \\
    Trend line & Fitted trend annotation. & Direct & \mbox{\texttt{createRegression}}, \mbox{\texttt{editRegression}} &  \\
    % Audit c0e47da: Derive mean from source data, then render a rule using the derived center and existing source scale.
    Average line & Line representing an average. & Partial &  & \mbox{\texttt{createIntervalData}} \(\rightarrow\) \mbox{\texttt{createRuleMark}} \(\rightarrow\) \mbox{\texttt{encodeY}} \\
    % Audit c0e47da: Create and populate a custom concrete graphic, not a dedicated embellishment facade.
    Embellishment & Additional decorative or explanatory graphics. & Partial &  & \mbox{\texttt{createGraphics}} \(\rightarrow\) \mbox{\texttt{editGraphics}} \\
    \midrule
    \rowcolor{white}\multicolumn{5}{@{}l@{}}{ Summary: 23 rows. Direct: 21 (91.3\%), Partial: 2 (8.7\%), Unsupported: 0 (0.0\%).} \\
    \bottomrule
  \end{tabularx}
\par\vspace{2pt}
  {\raggedright $\rightarrow$ denotes ordered calls, $+$ combined calls, and $/$ alternatives. See \autoref{sec:coverage-design} for the evaluation scope.\par}
\end{table*}

\begin{table*}[tp]
  \captionsetup{skip=3pt}
  \caption{\lib's coverage of the 24 operation categories in the taxonomy of Wang et al. \cite{wang23tvcg}. 22 operations are directly supported, and two require the composition of actions, i.e., are partially supported.}
  \label{tab:operation-coverage-1}
  \centering
  \renewcommand{\tabularxcolumn}[1]{m{#1}}
  \fontsize{5.6}{6.272}\selectfont
  \linespread{1}\selectfont
  \setlength{\tabcolsep}{1pt}
  \renewcommand{\arraystretch}{1.0}
  \setlength{\aboverulesep}{0.15ex}
  \setlength{\belowrulesep}{0.25ex}
  \rowcolors{2}{black!5}{white}
  \begin{tabularx}{\textwidth}{@{}c Z{0.4} Z{0.78} l Z{1.61} Z{1.21}@{}}
    \toprule
     & Operation & Definition & Coverage & Direct & Partial \\
    \midrule
    \cellcolor{white}\zsavepos{group-1-start} & Filter data & Retain records matching a condition. & Direct & \mbox{\texttt{filterData}}, \mbox{\texttt{filterMarks}} &  \\
    \cellcolor{white} & Aggregate data & Summarize data records. & Direct & \mbox{\texttt{encodeX}}, \mbox{\texttt{encodeY}}, \mbox{\texttt{createIntervalData}}, \mbox{\texttt{createSummaryData}} &  \\
    \cellcolor{white} & Bin data & Group values into intervals. & Direct & \mbox{\texttt{encodeHistogram}}, \mbox{\texttt{createHistogram}}, \mbox{\texttt{createBinData}}, \mbox{\texttt{createBin2DData}}, \mbox{\texttt{editBin2DData}} &  \\
    \cellcolor{white} & Set time unit & Set temporal granularity. & Direct & \mbox{\texttt{createTimeUnitData}} &  \\
    \cellcolor{white}\zsavepos{group-1-end}\smash{\raisebox{\dimexpr(\zposy{group-1-start}sp-\zposy{group-1-end}sp)/2-.5\height\relax}{\rotatebox{90}{Data}}} & Sort data & Order displayed values or categories. & Direct & \mbox{\texttt{orderCategories}}, \mbox{\texttt{removeCategoryOrder}} &  \\
    \cmidrule(lr){2-6}
    % Audit c0e47da: Replace the targeted existing mark with a chart facade, reusing raw data and compatible or explicitly distinct scales.
    \cellcolor{white}\zsavepos{group-2-start} & Set chart type & Assign or change chart type. & Partial &  & (\mbox{\texttt{removeMark}} \(\rightarrow\) \mbox{\texttt{createScatterPlot}}) / (\mbox{\texttt{removeMark}} \(\rightarrow\) \mbox{\texttt{createLinePlot}})  \\
    \cellcolor{white}\zsavepos{group-2-end}\smash{\raisebox{\dimexpr(\zposy{group-2-start}sp-\zposy{group-2-end}sp)/2-.5\height\relax}{\rotatebox{90}{Encoding}}} & Bind data to channel & Map data fields to visual channels. & Direct & \mbox{\texttt{encodeX}}, \mbox{\texttt{encodeY}}, \mbox{\texttt{encodeColor}}, \mbox{\texttt{encodeGroup}}, \mbox{\texttt{encodeShape}}, \mbox{\texttt{encodeSize}} &  \\
    \cmidrule(lr){2-6}
    \cellcolor{white}\zsavepos{group-3-start} & Create mark & Add data-encoding marks or series. & Direct & \mbox{\texttt{createPointMark}}, \mbox{\texttt{createLineMark}}, \mbox{\texttt{createBarMark}}, \mbox{\texttt{createAreaMark}}, \mbox{\texttt{createArcMark}}, \mbox{\texttt{createTextMark}} &  \\
    \cellcolor{white} & Set mark shape & Set data-mark shape or icon. & Direct & \mbox{\texttt{editPointMark}}, \mbox{\texttt{encodeShape}} &  \\
    \cellcolor{white}\zsavepos{group-3-end}\smash{\raisebox{\dimexpr(\zposy{group-3-start}sp-\zposy{group-3-end}sp)/2-.5\height\relax}{\rotatebox{90}{Mark}}} & Repeat by field & Repeat a visual unit by field. & Direct & \mbox{\texttt{facet}}, \mbox{\texttt{facetGrid}}, \mbox{\texttt{repeatCharts}} &  \\
    \cmidrule(lr){2-6}
    \cellcolor{white}\zsavepos{group-4-start} & Set color & Change element color. & Direct & \mbox{\texttt{editPointMark}}, \mbox{\texttt{editLineMark}}, \mbox{\texttt{editBarMark}}, \mbox{\texttt{editAreaMark}}, \mbox{\texttt{editCanvas}}, \mbox{\texttt{editTitle}} &  \\
    \cellcolor{white} & Set opacity & Change element transparency. & Direct & \mbox{\texttt{encodeOpacity}}, \mbox{\texttt{editPointMark}}, \mbox{\texttt{editLineMark}}, \mbox{\texttt{editBarMark}}, \mbox{\texttt{editAreaMark}}, \mbox{\texttt{highlightMarks}} &  \\
    \cellcolor{white} & Set size & Change element dimensions. & Direct & \mbox{\texttt{editCanvas}}, \mbox{\texttt{encodePointRadius}}, \mbox{\texttt{encodeBarWidth}}, \mbox{\texttt{editLineMark}}, \mbox{\texttt{editTickMark}}, \mbox{\texttt{editTextMark}} &  \\
    % Audit c0e47da: Construct a custom non-data graphic and populate its concrete shape.
    \cellcolor{white} & Set graphical shape or icon & Set non-data graphical shape or icon. & Partial &  & \mbox{\texttt{createGraphics}} \(\rightarrow\) \mbox{\texttt{editGraphics}} \\
    \cellcolor{white} & Set stroke & Change line or border appearance. & Direct & \mbox{\texttt{editPointMark}}, \mbox{\texttt{editLineMark}}, \mbox{\texttt{editBarMark}}, \mbox{\texttt{editAreaMark}}, \mbox{\texttt{editRectMark}}, \mbox{\texttt{editArcMark}} &  \\
    \cellcolor{white} & Set text font & Change text font properties. & Direct & \mbox{\texttt{editTextMark}}, \mbox{\texttt{editTitle}}, \mbox{\texttt{editLegendLabels}}, \mbox{\texttt{editLegendTitle}}, \mbox{\texttt{editXAxisLabels}}, \mbox{\texttt{editYAxisLabels}} &  \\
    \cellcolor{white} & Set text content & Change textual content. & Direct & \mbox{\texttt{encodeText}}, \mbox{\texttt{editTitle}}, \mbox{\texttt{editLegendTitle}}, \mbox{\texttt{editXAxisTitle}}, \mbox{\texttt{editYAxisTitle}} &  \\
    \cellcolor{white} & Add component & Add a chart component. & Direct & \mbox{\texttt{createTitle}}, \mbox{\texttt{createGrid}}, \mbox{\texttt{createXAxis}}, \mbox{\texttt{createYAxis}}, \mbox{\texttt{createLegend}}, \mbox{\texttt{createGuides}} &  \\
    \cellcolor{white}\zsavepos{group-4-end}\smash{\raisebox{\dimexpr(\zposy{group-4-start}sp-\zposy{group-4-end}sp)/2-.5\height\relax}{\rotatebox{90}{Styling}}} & Remove component & Remove a chart component. & Direct & \mbox{\texttt{removeTitle}}, \mbox{\texttt{removeGrid}}, \mbox{\texttt{removeXAxis}}, \mbox{\texttt{removeYAxis}}, \mbox{\texttt{removeLegend}}, \mbox{\texttt{removeThetaAxis}} &  \\
    \cmidrule(lr){2-6}
    \cellcolor{white}\zsavepos{group-5-start} & Place or arrange objects & Position or arrange chart elements. & Direct & \mbox{\texttt{editLegendLayout}}, \mbox{\texttt{editTitle}}, \mbox{\texttt{editXAxis}}, \mbox{\texttt{editYAxis}}, \mbox{\texttt{editCompositionLayout}}, \mbox{\texttt{facet}} &  \\
    % Audit c0e47da: Dedicated movement/offset support for text, title, or legend. highlightMarks is not geometric movement.
    \cellcolor{white} & Move object & Move an element by direction or offset. & Direct & \mbox{\texttt{editTextMark}}, \mbox{\texttt{editTitle}}, \mbox{\texttt{editLegendLayout}} &  \\
    \cellcolor{white}\zsavepos{group-5-end}\smash{\raisebox{\dimexpr(\zposy{group-5-start}sp-\zposy{group-5-end}sp)/2-.5\height\relax}{\rotatebox{90}{Layout}}} & Set offset or margin & Adjust spacing or offsets. & Direct & \mbox{\texttt{editCanvas}}, \mbox{\texttt{editLegendLayout}}, \mbox{\texttt{editTextMark}}, \mbox{\texttt{editFacetHeaders}}, \mbox{\texttt{editCompositionLayout}} &  \\
    \cmidrule(lr){2-6}
    \cellcolor{white}\zsavepos{group-6-start} & Add annotation & Add an explanatory annotation. & Direct & \mbox{\texttt{createAnnotation}}, \mbox{\texttt{createMarkLabels}}, \mbox{\texttt{createReferenceLine}}, \mbox{\texttt{createReferenceBand}} &  \\
    % Audit c0e47da: Remove an independently owned annotation mark; source-owned labels follow the source owner lifecycle.
    \cellcolor{white}\zsavepos{group-6-end}\smash{\raisebox{\dimexpr(\zposy{group-6-start}sp-\zposy{group-6-end}sp)/2-.5\height\relax}{\rotatebox{90}{Annotate}}} & Remove annotation & Remove an explanatory annotation. & Direct & \mbox{\texttt{removeMark}} &  \\
    \midrule
    \rowcolor{white}\multicolumn{6}{@{}l@{}}{ Summary: 24 rows. Direct: 22 (91.7\%), Partial: 2 (8.3\%), Unsupported: 0 (0.0\%).} \\
    \bottomrule
  \end{tabularx}
\par\vspace{2pt}
  {\raggedright $\rightarrow$ denotes ordered calls, $+$ combined calls, and $/$ alternatives. See \autoref{sec:coverage-design} for the evaluation scope.\par}
\end{table*}

\begin{table*}[tp]
  \captionsetup{skip=3pt}
  \caption{\lib's coverage of 50 VisTalk intent entries from Wang et al. \cite{wang23tvcg}: 43 directly supported, five supported through composition, and two unsupported.}
  \label{tab:intent-coverage}
  \centering
  \renewcommand{\tabularxcolumn}[1]{m{#1}}
  \fontsize{5.6}{6.272}\selectfont
  \linespread{1}\selectfont
  \setlength{\tabcolsep}{1pt}
  \renewcommand{\arraystretch}{1.0}
  \setlength{\aboverulesep}{0.15ex}
  \setlength{\belowrulesep}{0.25ex}
  \rowcolors{2}{black!5}{white}
  \begin{tabularx}{\textwidth}{@{}c Z{0.7} Z{0.63} Z{0.34} l Z{1.79} Z{1.54}@{}}
    \toprule
     & Intent & Operation & Object & Coverage & Direct & Partial \\
    \midrule
    \cellcolor{white}\zsavepos{group-7-start} & Change\allowbreak Aggregation & Aggregate data & Data field & Direct & \mbox{\texttt{encodeX}}, \mbox{\texttt{encodeY}} &  \\
    \cellcolor{white} & Filter & Filter data & Data point & Direct & \mbox{\texttt{filterMarks}} &  \\
    % Audit c0e47da: Automatic anomaly identification plus removal on an existing chart. Supplied threshold filtering alone is not detection.
    \cellcolor{white} & Remove\allowbreak Anomaly\allowbreak Data\allowbreak Points & Filter data & Data point & Unsupported &  &  \\
    % Audit c0e47da: Blank values: absent/null, empty string, or literal (blank). Sequential filters use mode compose. No whitespace trimming.
    \cellcolor{white} & Remove\allowbreak Blank\allowbreak Data\allowbreak Points & Filter data & Data point & Partial &  & \mbox{\texttt{filterMarks}} \(\rightarrow\) \mbox{\texttt{removeMarks}} \\
    \cellcolor{white} & Remove\allowbreak Data\allowbreak Points & Filter data & Data point & Direct & \mbox{\texttt{filterMarks}} &  \\
    % Audit c0e47da: removeMark for an independent series layer; filterMarks for a series identified by a supported field selector.
    \cellcolor{white} & Remove\allowbreak Series & Filter data & Data point & Direct & \mbox{\texttt{removeMark}}, \mbox{\texttt{filterMarks}} &  \\
    \cellcolor{white} & Filter & Filter data & Data range & Direct & \mbox{\texttt{filterData}}, \mbox{\texttt{filterMarks}} &  \\
    \cellcolor{white} & Remove\allowbreak Data\allowbreak Points & Filter data & Data range & Direct & \mbox{\texttt{filterMarks}} &  \\
    \cellcolor{white} & Sort\allowbreak Asc & Sort data & Data field & Direct & \mbox{\texttt{orderCategories}} &  \\
    \cellcolor{white} & Sort\allowbreak Default & Sort data & Data field & Direct & \mbox{\texttt{removeCategoryOrder}} &  \\
    \cellcolor{white}\zsavepos{group-7-end}\smash{\raisebox{\dimexpr(\zposy{group-7-start}sp-\zposy{group-7-end}sp)/2-.5\height\relax}{\rotatebox{90}{Data}}} & Sort\allowbreak Desc & Sort data & Data field & Direct & \mbox{\texttt{orderCategories}} &  \\
    \cmidrule(lr){2-7}
    \cellcolor{white}\zsavepos{group-8-start} & Bind\allowbreak X & Bind data to channel & Axis & Direct & \mbox{\texttt{encodeX}} &  \\
    \cellcolor{white} & Bind\allowbreak Y & Bind data to channel & Axis & Direct & \mbox{\texttt{encodeY}} &  \\
    % Audit c0e47da: Both position encodings must be reassigned using compatible scale definitions.
    \cellcolor{white} & Swap\allowbreak Axis & Bind data to channel & Axis & Partial &  & \mbox{\texttt{encodeX}} + \mbox{\texttt{encodeY}} \\
    \cellcolor{white} & Bind\allowbreak Series & Bind data to channel & Mark channel & Direct & \mbox{\texttt{encodeColor}}, \mbox{\texttt{encodeGroup}}, \mbox{\texttt{encodeShape}}, \mbox{\texttt{encodeSize}}, \mbox{\texttt{encodeOpacity}}, \mbox{\texttt{encodeStrokeDash}} &  \\
    % Audit c0e47da: Replace, rather than layer over, the targeted existing chart mark.
    \cellcolor{white}\zsavepos{group-8-end}\smash{\raisebox{\dimexpr(\zposy{group-8-start}sp-\zposy{group-8-end}sp)/2-.5\height\relax}{\rotatebox{90}{Encoding}}} & Set\allowbreak Chart\allowbreak Type & Set chart type & Chart & Partial &  & (\mbox{\texttt{removeMark}} \(\rightarrow\) \mbox{\texttt{createScatterPlot}}) / (\mbox{\texttt{removeMark}} \(\rightarrow\) \mbox{\texttt{createLinePlot}})  \\
    \cmidrule(lr){2-7}
    % Audit c0e47da: A chart facade adds another fully encoded series layer to existing data. Verified with createLinePlot while original layer remains.
    \cellcolor{white}\zsavepos{group-9-start} & Add\allowbreak Series & Create mark & Mark & Direct & \mbox{\texttt{createLinePlot}}, \mbox{\texttt{createScatterPlot}}, \mbox{\texttt{createBarPlot}} &  \\
    \cellcolor{white} & Repeat & Repeat by field & Chart & Direct & \mbox{\texttt{facet}} &  \\
    \cellcolor{white}\zsavepos{group-9-end}\smash{\raisebox{\dimexpr(\zposy{group-9-start}sp-\zposy{group-9-end}sp)/2-.5\height\relax}{\rotatebox{90}{Mark}}} & Set\allowbreak Mark & Set mark shape & Mark & Direct & \mbox{\texttt{editPointMark}}, \mbox{\texttt{encodeShape}} &  \\
    \cmidrule(lr){2-7}
    \cellcolor{white}\zsavepos{group-10-start} & Add\allowbreak Grid\allowbreak Lines & Add component & Gridline & Direct & \mbox{\texttt{createGrid}}, \mbox{\texttt{createHorizontalGrid}}, \mbox{\texttt{createVerticalGrid}} &  \\
    % Audit c0e47da: Create series encoding first, then its legend; prior series binding is not assumed for this compound intent.
    \cellcolor{white} & Bind\allowbreak Series & Add component & Legend & Partial &  & (\mbox{\texttt{encodeColor}} \(\rightarrow\) \mbox{\texttt{createLegend}}) / (\mbox{\texttt{encodeGroup}} \(\rightarrow\) \mbox{\texttt{createLegend}}) \\
    \cellcolor{white} & Add\allowbreak Chart\allowbreak Title & Add component & Title & Direct & \mbox{\texttt{createTitle}} &  \\
    \cellcolor{white} & Remove\allowbreak YAxis & Remove component & Axis & Direct & \mbox{\texttt{removeYAxis}} &  \\
    \cellcolor{white} & Remove\allowbreak Grid\allowbreak Lines & Remove component & Gridline & Direct & \mbox{\texttt{removeGrid}} &  \\
    \cellcolor{white} & Remove\allowbreak Chart\allowbreak Title & Remove component & Title & Direct & \mbox{\texttt{removeTitle}} &  \\
    \cellcolor{white} & Set\allowbreak Color & Set color & Canvas & Direct & \mbox{\texttt{editCanvas}} &  \\
    \cellcolor{white} & Highlight & Set color & Mark & Direct & \mbox{\texttt{highlightMarks}} &  \\
    \cellcolor{white} & Highlight\allowbreak Negative & Set color & Mark & Direct & \mbox{\texttt{highlightMarks}} &  \\
    % Audit c0e47da: Automatic outlier identification and highlighting on an existing chart. highlightMarks with a supplied selector is ordinary highlighting, not detection.
    \cellcolor{white} & Highlight\allowbreak Outlier & Set color & Mark & Unsupported &  &  \\
    \cellcolor{white} & Set\allowbreak Color & Set color & Mark & Direct & \mbox{\texttt{editPointMark}}, \mbox{\texttt{editLineMark}}, \mbox{\texttt{editBarMark}}, \mbox{\texttt{editAreaMark}}, \mbox{\texttt{editRectMark}}, \mbox{\texttt{editArcMark}} &  \\
    \cellcolor{white} & Set\allowbreak Color & Set color & Title & Direct & \mbox{\texttt{editTitle}} &  \\
    \cellcolor{white} & Decrease\allowbreak Height & Set size & Chart & Direct & \mbox{\texttt{editCanvas}} &  \\
    \cellcolor{white} & Decrease\allowbreak Width & Set size & Chart & Direct & \mbox{\texttt{editCanvas}} &  \\
    \cellcolor{white} & Increase\allowbreak Height & Set size & Chart & Direct & \mbox{\texttt{editCanvas}} &  \\
    \cellcolor{white} & Increase\allowbreak Width & Set size & Chart & Direct & \mbox{\texttt{editCanvas}} &  \\
    \cellcolor{white} & Decrease\allowbreak Data\allowbreak Point\allowbreak Width & Set size & Mark & Direct & \mbox{\texttt{encodePointRadius}}, \mbox{\texttt{encodeBarWidth}}, \mbox{\texttt{editLineMark}}, \mbox{\texttt{editTickMark}}, \mbox{\texttt{editTextMark}} &  \\
    \cellcolor{white} & Increase\allowbreak Data\allowbreak Point\allowbreak Width & Set size & Mark & Direct & \mbox{\texttt{encodePointRadius}}, \mbox{\texttt{encodeBarWidth}}, \mbox{\texttt{editLineMark}}, \mbox{\texttt{editTickMark}}, \mbox{\texttt{editTextMark}} &  \\
    \cellcolor{white}\zsavepos{group-10-end}\smash{\raisebox{\dimexpr(\zposy{group-10-start}sp-\zposy{group-10-end}sp)/2-.5\height\relax}{\rotatebox{90}{Styling}}} & Add\allowbreak Chart\allowbreak Title & Set text content & Title & Direct & \mbox{\texttt{editTitle}} &  \\
    \cmidrule(lr){2-7}
    \cellcolor{white}\zsavepos{group-11-start} & Change\allowbreak Position & Place or arrange objects & Axis & Direct & \mbox{\texttt{editXAxis}}, \mbox{\texttt{editYAxis}}, \mbox{\texttt{editRadialAxis}} &  \\
    % Audit c0e47da: columns applies to an existing facet composition, not arbitrary non-facet layouts.
    \cellcolor{white} & Show\allowbreak Columns & Place or arrange objects & Chart & Direct & \mbox{\texttt{editCompositionLayout}} &  \\
    \cellcolor{white} & Change\allowbreak Layout & Place or arrange objects & Legend & Direct & \mbox{\texttt{editLegendLayout}}, \mbox{\texttt{editLegend}} &  \\
    \cellcolor{white} & Change\allowbreak Position & Place or arrange objects & Legend & Direct & \mbox{\texttt{editLegendLayout}}, \mbox{\texttt{editLegend}} &  \\
    \cellcolor{white} & Change\allowbreak Position & Place or arrange objects & Title & Direct & \mbox{\texttt{editTitle}} &  \\
    \cellcolor{white}\zsavepos{group-11-end}\smash{\raisebox{\dimexpr(\zposy{group-11-start}sp-\zposy{group-11-end}sp)/2-.5\height\relax}{\rotatebox{90}{Layout}}} & Increase\allowbreak Margin & Set offset or margin & Chart & Direct & \mbox{\texttt{editCanvas}} &  \\
    \cmidrule(lr){2-7}
    % Audit c0e47da: Derive mean from source data, create a rule, then encode the derived center using the original source scale.
    \cellcolor{white}\zsavepos{group-12-start} & Add\allowbreak Line & Add annotation & Average line & Partial &  & \mbox{\texttt{createIntervalData}} \(\rightarrow\) \mbox{\texttt{createRuleMark}} \(\rightarrow\) \mbox{\texttt{encodeY}} \\
    \cellcolor{white} & Add\allowbreak Data\allowbreak Labels & Add annotation & Label & Direct & \mbox{\texttt{createMarkLabels}} &  \\
    \cellcolor{white} & Add\allowbreak Series\allowbreak Labels & Add annotation & Label & Direct & \mbox{\texttt{createMarkLabels}} &  \\
    \cellcolor{white} & Add\allowbreak Rectangle & Add annotation & Reference band & Direct & \mbox{\texttt{createReferenceBand}} &  \\
    \cellcolor{white} & Add\allowbreak Line & Add annotation & Reference line & Direct & \mbox{\texttt{createReferenceLine}} &  \\
    \cellcolor{white}\zsavepos{group-12-end}\smash{\raisebox{\dimexpr(\zposy{group-12-start}sp-\zposy{group-12-end}sp)/2-.5\height\relax}{\rotatebox{90}{Annotate}}} & Add\allowbreak Trend\allowbreak Line & Add annotation & Trend line & Direct & \mbox{\texttt{createRegression}} &  \\
    \midrule
    \rowcolor{white}\multicolumn{7}{@{}l@{}}{ Summary: 50 rows. Direct: 43 (86.0\%), Partial: 5 (10.0\%), Unsupported: 2 (4.0\%).} \\
    \bottomrule
  \end{tabularx}
\par\vspace{2pt}
  {\raggedright  $\rightarrow$ denotes ordered calls, $+$ combined calls, and $/$ alternatives. See \autoref{sec:coverage-design} for the evaluation scope.\par}
\end{table*}

\subsection{Objectives and Design}
\label{sec:coverage-design}
% Repository audit snapshot: full public API, ggaction main c0e47da6e213852213bcb04eb19031a1a6a63cd7 (package version 0.0.13).

We aim to assess how well \lib covers the chart authoring intents that designers commonly express in practice. 
To this end, we adopt the collection of intents from Wang et al. \cite{wang23tvcg}, which is derived from user studies of natural language-based chart authoring. We evaluate a fixed snapshot of the API of \lib (v0.0.13) by mapping taxonomy items to implemented public actions. 
The detailed evaluation procedure is as follows:

\myparagraph{Taxonomy extraction}
Following Wang et al. \cite{wang23tvcg}, we extract the taxonomy along two axes: \textit{objects} and \textit{operations}, as well as their \textit{intents}, which are the combinations of objects and operations. 
These axes correspond to the nouns and verbs, respectively, in our action design (\autoref{sec:grammarmodel}).
Although we initially drew on Wang et al. when designing our verb--noun action schema, we did not consult their intent taxonomy to determine which intents \lib{} should support. Thus, at least for the version of \lib{} evaluated here, its action vocabulary was developed independently of the intent taxonomy used in our coverage analysis.
We conduct qualitative coding with two coders to extract the objects and operations. 
One is an author of this paper, and the other is an external researcher not involved in the project. 
The external coder has prior high-level familiarity with \lib through discussions with the authors but is not familiar with its implementation details. 
Both coders hold Ph.D. degrees and conduct research primarily in data visualization.
The coders independently review Section 4 (``Editing Actions'') of the reference paper, together with the associated Figure 2 and Table 1. 
After independent coding, we merge the results by taking the union of the two code sets to reduce the risk of omitting a relevant taxonomy item, i.e., false negatives. Cohen's $\kappa$ is 0.79.
For the combinations of operations and objects (which correspond to individual actions in our system), we use the atomic editing intents in VisTalk, an authoring tool introduced in the reference paper \cite{wang23tvcg}. We refer to the original source code\footnote{\url{https://github.com/microsoft/VisTalk}} to do so.

\myparagraph{Investigating the coverage}
We collaborate with an LLM to investigate the coverage. 
For each operation, object, and intent, we prompt an LLM to examine action calls needed to perform an edit, assuming that the raw data and a compatible chart context already exist. We use the Codex application to call GPT-6 Astra with ``very high'' reasoning (detailed prompts in Appendix A).
Then, we classify an item as \textit{directly supported} when a dedicated action implements the item. An entry is \textit{partially supported} when it requires a composition of two or three action calls. 
We classify an item as \textit{unsupported} when no mapping is found within a budget of four calls.
Finally, a single coder, an author of the paper, reviews the results and validates the mappings.
The coder revises mappings proposed by the LLM, allowing the LLM to maximize recall while human review ensures precision. 

\subsection{Results and Discussion}

We discuss the results of our investigation. In summary, \lib provides broad coverage of the authoring intent taxonomy. At the same time, our analysis reveals several remaining gaps, highlighting opportunities for future extensions. 
The detailed coverage analysis results, including the mapped actions from the \lib library, are in \autoref{tab:object-coverage-1}, \ref{tab:operation-coverage-1}, and \ref{tab:intent-coverage}.

\myparagraph{Quantitative results} Across the 97 entries, we find a total of 86 direct (88.7\%), nine partial (9.3\%), and two unsupported (2.1\%) mappings. These results demonstrate broad coverage of \lib on common authoring intents in practice.

\myparagraph{Qualitative analysis of the coverage}
We examine the partial and unsupported entries to identify current weaknesses of \lib.
First, some higher-level intents are not directly captured by its current action vocabulary.
For example, swapping axes must be expressed by reassigning both positional channels through \codeverb{encodeX} and \codeverb{encodeY}, rather than as a single high-level action.
This suggests an opportunity to expand the vocabulary by identifying common high-level actions through observations of real-world authoring practices.

Second, \lib currently lacks direct support for intents that require identifying analytically meaningful subsets of the data, such as highlighting anomalies or outliers.
Supporting such intents would require integrating data transformation and analysis capabilities, including not only outlier detection but also clustering or dimensionality reduction.
While these capabilities may broaden the range of supported authoring intents, they would also increase the complexity of the grammar and its implementation.
We discuss this tradeoff further in \autoref{sec:tradeoff}.

\section{Evaluating Machine Interpretability}

\label{sec:machine}

We discuss our experiment investigating the machine interpretability of \lib and other competitor grammars.

% examining how well small language models can interpret \lib specifications.

\subsection{Objectives and Design}

We aim to compare the machine interpretability of \lib and competing grammars, i.e., the extent to which their semantics and syntax can be readily learned and generated by language models. 
Specifically, we examine how effectively SLMs can learn to produce valid grammar specifications. Our rationale is that better performance with smaller models indicates that a grammar enables machines to translate users' authoring intent into rendered graphics with lower computational and monetary cost (in the case of calling an API from commercial services).
In practical terms, higher machine interpretability means that effective authoring support can be built with fewer computational resources, thereby reducing deployment costs.
We do not evaluate LLMs because their large capacity may obscure differences in grammar interpretability: given sufficient context, documentation, and reasoning time, frontier LLMs may perform well across all grammars regardless of their inherent interpretability. Moreover, LLMs may have been exposed to existing visualization grammars during pretraining, introducing an additional bias into the comparison.

We first construct pairs of natural language prompts and corresponding ground truth code specifications, and then fine-tune SLMs to generate the specifications from the prompts. 
We use the accuracy with which the fine-tuned models reproduce the ground truth specifications as a proxy for machine interpretability. The detailed experimental setup is as follows:

\myparagraph{Chart grammars}
We compare \lib{} with Vega-Lite, ggplot2, and Plotly, three representative high-level visualization authoring grammars that, like \lib{}, expose chart semantics rather than low-level rendering primitives. 
We select these systems because they are widely adopted in practice; as of September 2026, their official GitHub repositories have approximately 5.5K, 7.0K, and 18.8K stars, respectively. Prior work also describes these three grammars as well known and widely used \cite{pu23chi, stoiber24vi}.

% ggplot2 and Vega-Lite as popular visualization grammars with substantial practitioner and research ecosystems \cite{pu2023grammar}, while Plotly has been described as a widely used high-level declarative charting library \cite{stoiber2024visahoi}.

\myparagraph{Training and testing dataset}
We aim to simulate a common scenario in which chart designers progressively refine and accumulate their design intent through a sequence of prompts \cite{kapadnis2026arxiv, jeon26tvcg, zhao2025acl}.
To this end, we want to construct a dataset of input--output pairs, where each input consists of a current chart specification and a natural language prompt describing a desired revision, and the output is the revised chart specification.
However, existing datasets of this form rely largely on synthetic charts or natural language utterances describing simple revisions of charts \cite{kapadnis2026arxiv, shao20acl}, making them less appropriate for capturing the dynamics of real-world chart authoring.
We thus use real-world chart specifications to create a dataset, leveraging a collection constructed by Ko et al. 
\cite{ko24chi}. This set consists of 1,981 Vega-Lite charts collected from the web, where the diversity in terms of chart type, complexity, and chart composition is higher than other collections of chart specifications \cite{kim20chi, zhao22tvcg}.

To construct a dataset for training and evaluating SLMs, we first randomly sample 1,000 charts from the collection, considering our computational budget. We then generate natural language descriptions of these charts as the input component of the dataset, following the captioning pipeline proposed by Ko et al. \cite{ko24chi}. Specifically, we generate L1 captions, which describe the elemental visual and graphical characteristics of a chart. We focus on this level because, during chart authoring, users commonly express how a chart should appear rather than the higher-level takeaways it should convey \cite{srinivasan21chi}.
To prevent the generated descriptions from being biased toward the structure or terminology of Vega-Lite specifications, we first render each specification as an image and provide only the rendered chart to a vision-language model (VLM) for caption generation. We use the most capable VLM available to us at the time of data construction, GPT-5.6 Sol, configured with ``xhigh'' reasoning effort.
We instruct the VLM to produce a sufficiently detailed description such that the chart could, in principle, be faithfully reconstructed from the description alone. Each description consists of seven sentences, ordered from the most salient structural decisions, such as chart type and overall composition, to increasingly fine-grained and stylistic details. 

Finally, we use these seven sentences to construct a seven-turn chart-authoring trajectory for each chart and each visualization grammar under evaluation. At the first turn, we provide only the first sentence and ask the LLM (again, GPT-5.6 Sol with very high reasoning) to generate an initial specification in the target grammar. At each subsequent turn, we provide the current specification together with the next sentence and ask the model to generate a patch that incorporates the newly introduced design intent. We then apply the patch to obtain the specification used as input to the following turn. We repeat this procedure for all seven sentences and for every competing grammar. Consequently, each chart--grammar pair yields a sequence of seven input--output examples that progressively reconstructs the target chart. 
Note that we use an OpenAI API call to prompt LLMs. We report all prompts used to generate our dataset in Appendix B. 

\myparagraph{SLM settings}
We select the Qwen2.5-Coder Instruct \cite{hui2024qwen25} series as our primary SLM family. We choose this family because it is specifically pretrained for code understanding and generation, making it well aligned with the target task of our experiment. In addition, the models do not incorporate auxiliary capabilities such as dedicated reasoning mechanisms or vision inputs, reducing the need to control for additional factors. Finally, the series provides models across several relatively small and systematically varying sizes, allowing us to examine the effect of model capacity while keeping the model architecture and training family constant.

We conduct three experiments with SLMs to examine the interpretability of grammars:

\myparagraphit{Experiment 1: General performance and the effect of model size}
We fine-tune the 0.5B, 1.5B, and 3B variants of Qwen2.5-Coder Instruct (4-bit quantized) on the full dataset to evaluate their overall ability to learn visualization grammars and to examine how machine interpretability varies with model size. We randomly split the dataset into 800 training, 100 validation, and 100 test examples, using the same split across all grammars.

\myparagraphit{Experiment 2: Effect of data size}
We also want to investigate how model performance varies with data size. We fine-tune the 0.5B model with 100, 200, 400, and 800 chart samples, where we use the same validation and test examples. 

\vspace{3pt}
In addition, we examine whether the results of the above two experiments generalize to other model families through the following experiment: 

\myparagraphit{Experiment 3: Generalization across different models}
We fine-tune Llama 3.2 1B Instruct \cite{meta2024llama32} and Gemma 3 1B IT \cite{gemmateam2025gemma3t} with the full training dataset and examine how the results compare with those of Experiment 1.

\begin{figure*}[t]
    \centering
    \includegraphics[width=\textwidth]{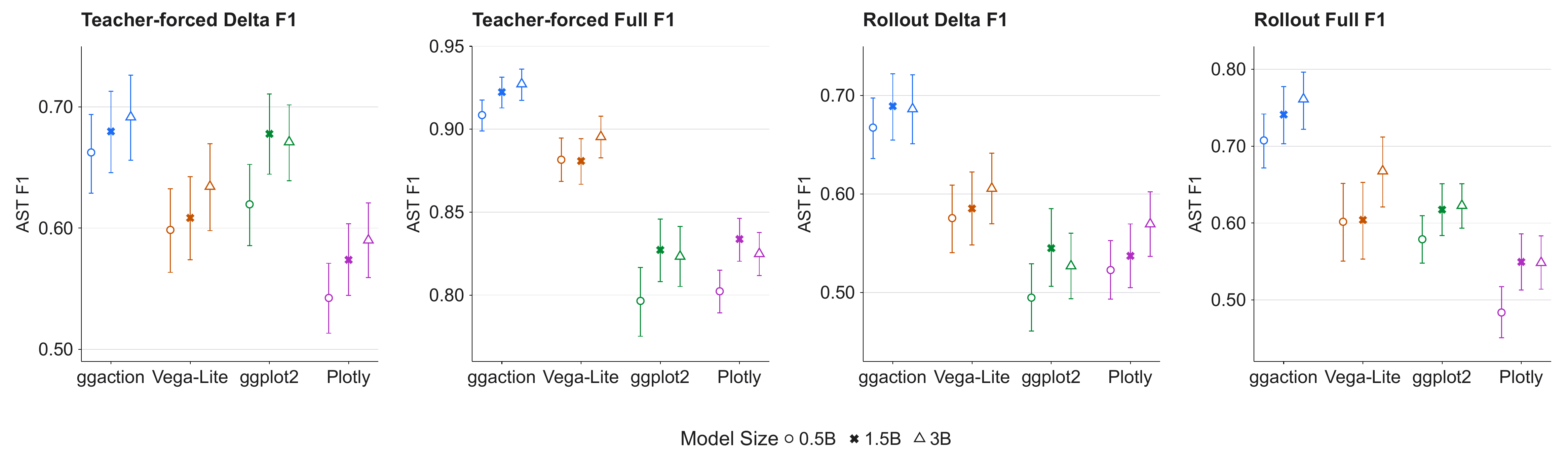}
    \caption{The performance of fine-tuned SLMs (Qwen2.5-Coder Instruct) with different sizes in generating chart specifications from natural language prompts. In summary, models showed better performance in understanding and generating \lib code.
    Error bars indicate 95\% confidence intervals.}
    \label{fig:machine-exp1}
\end{figure*}

\myparagraph{Fine-tuning settings}
For all settings, we perform LoRA fine-tuning with a batch size of 4, a maximum context length of 4,096 tokens, and AdamW optimization. Models are trained for up to three epochs, with early stopping if the validation loss does not improve over three consecutive evaluations. Each validation is done once for 100 charts.

\myparagraph{Apparatus}
We conduct all fine-tuning experiments on an Apple Mac Mini M4 with 16 GB of RAM. Our selection of SLMs also reflects this hardware constraint, as all chosen models can be feasibly fine-tuned with LoRA on this setup.

\myparagraph{Measurements}
We evaluate model accuracy as a proxy for machine interpretability.
As noted above, the model generates a patch for each turn, which we apply to the current specification to obtain a revised specification.
We compare this result with the ground truth respecification.
As two equivalent specifications can differ in superficial ways, such as whitespace, key ordering, fonts, or title text, we do not use exact string matching (details in Appendix B).
Instead, we parse each specification \(C\) into an abstract syntax tree  (AST) \(A(C)\) while removing these differences.

We evaluate each prediction from two complementary perspectives.
First, \textit{Full F1} measures how similar the complete predicted specification is to the complete ground-truth specification.
Second, \textit{Delta F1} focuses only on the revision made at the current turn: it compares the AST nodes added or removed by the model with those that should have been added or removed according to the ground truth.

Formally, for turn \(t\),
\begin{equation}
\begin{aligned}
\operatorname{Full\_F1}_t
&= \operatorname{F1}
   \bigl(A(\hat{C}_t), A(C_t^*)\bigr), \\
\operatorname{Delta\_F1}_t
&= \operatorname{F1}
   \Bigl(
   \Delta\bigl(A(I_t), A(\hat{C}_t)\bigr),
   \Delta\bigl(A(C_{t-1}^*), A(C_t^*)\bigr)
   \Bigr),
\end{aligned}
\end{equation}
where \(\hat{C}_t\) is the model-generated specification at turn \(t\), \(C_t^*\) is the corresponding ground-truth specification, \(I_t\) is the specification provided to the model as input at that turn, and \(\Delta\) returns the multiset of AST-node additions and removals between two specifications.

A potential limitation of this metric is that an AST node does not necessarily represent the same amount of meaning across grammars. 
If the same intent can be expressed with one node in one grammar but requires several coordinated nodes in another, errors in the latter may receive a larger penalty. 
In this sense, the metric can disadvantage grammars with more structurally complex specifications. However, we view this additional structural complexity as part of the generation burden imposed by the grammar, rather than purely as a measurement artifact. Developing alternative metrics that better separate semantic correctness from representational complexity would be an interesting direction for future work.

We compute these metrics under two evaluation settings.
In \textit{teacher-forced evaluation}, the model always receives the correct specification from the previous turn, i.e., \(I_t=C_{t-1}^*\).
This isolates the model's ability to perform each individual revision without being affected by earlier mistakes.
In \textit{rollout evaluation}, the model instead receives its own output from the previous turn, i.e., \(I_t=\hat{C}_{t-1}\).
This evaluates the model over the full chart-authoring trajectory and captures how errors accumulate across turns.
We report the mean Full F1 and Delta F1 scores over the test charts, together with 95\% confidence intervals estimated by chart-level bootstrap resampling.

\subsection{Results}

The results can be summarized as follows:

\myparagraph{Experiment 1 (General results and model capacity)}
As shown in Fig.~\ref{fig:machine-exp1}, the models generally learned the representation of \lib more effectively than those of the competing grammars. Across model sizes, \lib achieved the highest F1 scores in most cases and across all evaluation metrics, followed overall by Vega-Lite, ggplot2, and Plotly. 
Performance improved consistently with model capacity regardless of the underlying grammar. 
The performance gap between \lib and the competing grammars was relatively modest under the teacher-forced setting, whereas it became substantially more pronounced under rollout. For example, the difference in F1 score between \lib and ggplot2 was comparatively small with teacher forcing but widened considerably during rollout.

\begin{figure*}[t]
    \centering
    \includegraphics[width=\textwidth]{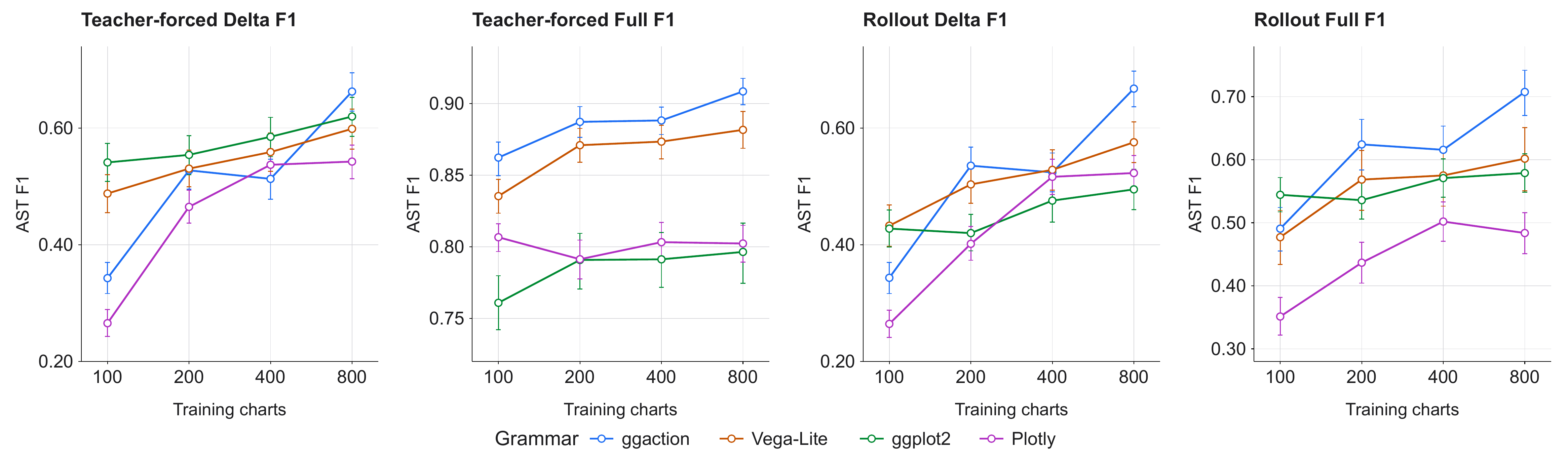}
    \caption{The trend in fine-tuned model performance as the size of the training data increases. With limited training data, \lib performs on par with or slightly worse than the competing grammars in some cases, but establishes a clear advantage as more training data becomes available. Error bars indicate 95\% confidence intervals.}
    \label{fig:machine-exp2}
\end{figure*}

\myparagraph{Experiment 2 (Data size)}
In summary (Fig.~\ref{fig:machine-exp2}), \lib shows mixed performance relative to the competing grammars when trained on small datasets, but establishes a consistent advantage as the training data increases. 
For example, with 100 training charts, the model trained on \lib shows a lower Delta F1 score than those trained on Vega-Lite and ggplot2, whereas it surpasses both when trained on the full dataset of 800 charts. 
Moreover, unlike the competing grammars, \lib continues to benefit substantially from additional training data, with no clear indication of performance saturation. 

\begin{figure*}[t]
    \centering
    \includegraphics[width=\textwidth]{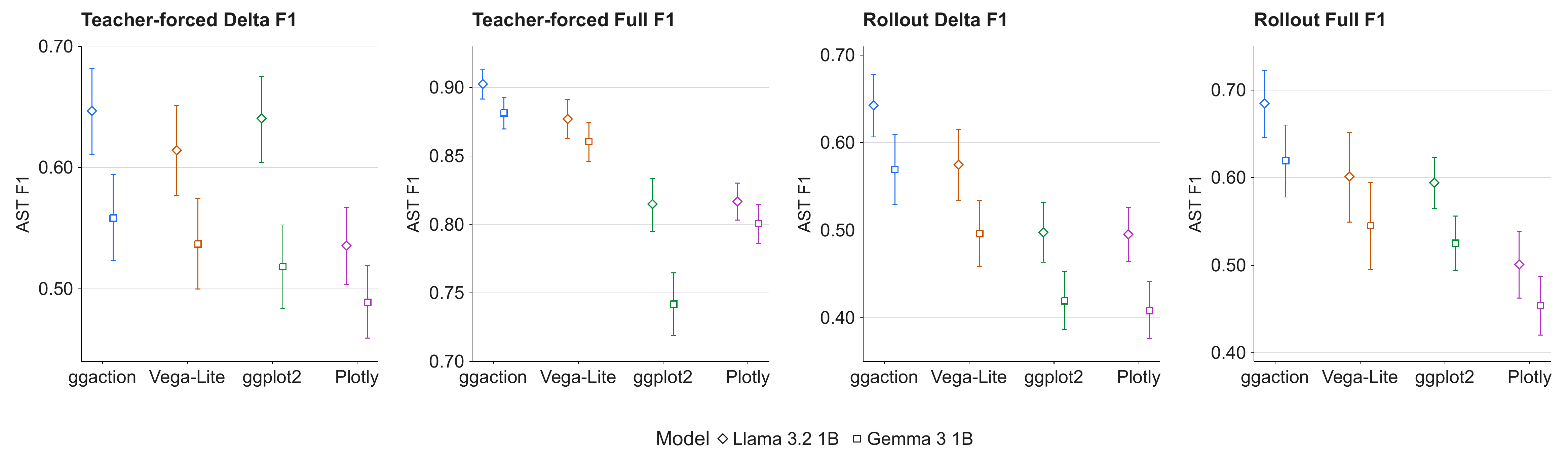}
    \caption{The reproduction of experiment 1 with two alternative models: Llama 3.2 1B Instruct and Gemma 3 1B IT. As with previous experiments, the models performed well in learning \lib compared to competitor grammars.}
    \label{fig:machine-exp3}
\end{figure*}

\myparagraph{Experiment 3 (Different models)}
As shown in Fig.~\ref{fig:machine-exp3}, the results with alternative SLMs are consistent with those of experiment 1: models fine-tuned on \lib generally outperform those fine-tuned on the competing grammars.

\subsection{Discussion}

The results suggest that \lib is comparatively easy for SLMs to learn and generate. 
Notably, the advantage of \lib becomes more pronounced under rollout evaluation, where models must rely on their own previous outputs. This suggests that the benefit of \lib lies not only in facilitating the generation of individual specifications but also in enabling more reliable iterative chart authoring.

Experiment 2 provides further evidence for this interpretation. Although \lib does not consistently outperform competing grammars with limited training data, this result may be due to pretraining exposure: competitor grammars are already well established and thus have been encountered by pretrained models, whereas \lib is entirely new to them. Despite this disadvantage, the performance of models trained on \lib continues to improve substantially as more examples become available and eventually surpasses the competing grammars, without showing a comparable degree of saturation. 
This suggests that \lib can be effectively learned from data. 
The consistent results across other models (experiment 3) further suggest that this behavior is not specific to a particular model family.

Overall, these results provide consistent evidence that \lib has comparatively high machine interpretability. In practical terms, this suggests that systems built on \lib may support effective chart authoring with relatively few computational resources, potentially enabling lightweight deployments (e.g., applications running only in the browser).

\section{Evaluating Human Interpretability}

\label{sec:humaninter}

We present our user study to evaluate the human interpretability of \lib compared with alternative grammars.

\subsection{Objectives and Design}

We evaluate how well novice chart designers interpret chart specifications written in different grammars.
This comparison is motivated by the premise that more interpretable code makes LLM-generated outputs easier to inspect and debug---an especially important property when using smaller or open-source models \cite{md26jss}.

We provide participants with code snippets written in different visualization grammars, including \lib, along with sample data, and check whether they can decode the visual representation from the code. This is done by asking participants to freely draw the visualization using colored pens and pencils based on the given code. We then ask data visualization experts to grade the drawings' quality based on how well they reflect the original code. The detailed experimental setup is as follows:

\myparagraph{Participants}
We recruit 16 participants (aged 24--56 years; \(M = 30.3\), \(SD = 7.6\); seven men and nine women) through advertisements posted on a local university community and snowball sampling \cite{goodman61ams}. 
This encompasses five graduate students and three postdoctoral researchers working in diverse disciplines including computer science, biochemistry and environmental engineering, and eight office workers. 
We require participants to have experience authoring charts by writing a program, either manually or with language model assistance, but with limited formal training in data visualization---for example, they work outside visualization research and, if students, have not taken courses in data visualization or visual analytics. 
These criteria reflect our target population: individuals who need to create visualizations but lack specialized expertise.

\myparagraph{Visualization grammars}
As with our machine experiment (\autoref{sec:machine}), we compare four visualization grammars: \lib, Vega-Lite, Plotly, and ggplot2. 

\myparagraph{Visualizations}
We design eight visualizations accompanied by sample data for participants to read specifications and draw the possible output. 
We first set three requirements that charts should satisfy:

\begin{itemize}[leftmargin=9pt]
\item \textbf{(R1) }\textit{Drawable:} The charts and corresponding data should be sufficiently simple for participants to reproduce by hand with reasonable fidelity.
\item \textbf{(R2) }\textit{Representative:} The charts should cover chart types and visual representations commonly encountered in practical chart authoring.
\item \textbf{(R3) }\textit{Equivalence:} The same underlying visualization should be expressible in all four grammars, without relying on features that are unique in a particular grammar.
\end{itemize}

\begin{figure*}[ht!]
    \centering
    \includegraphics[width=0.912\linewidth]{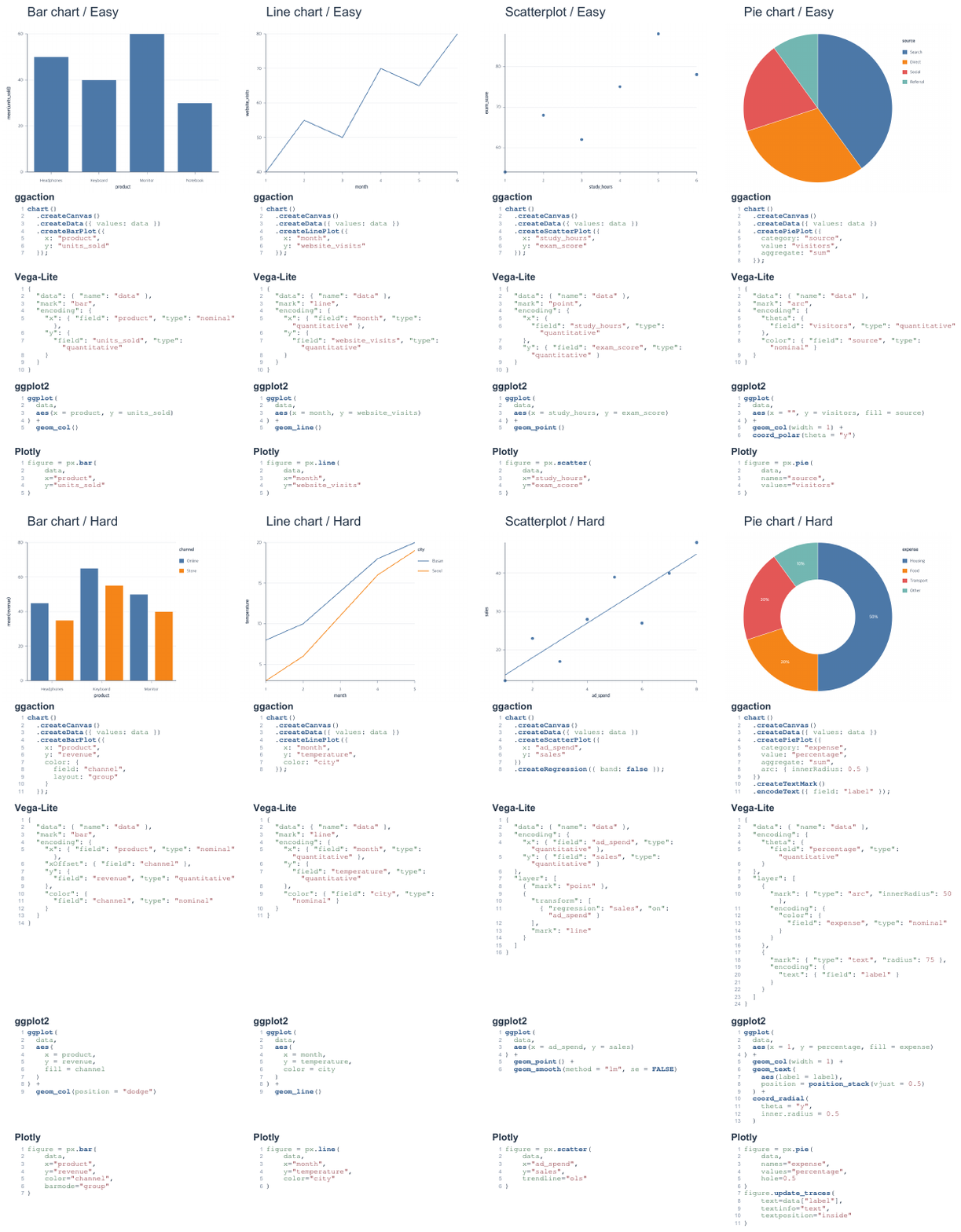}
    \vspace{-6mm}
    \caption{The charts that participants are asked to reproduce by hand in our human experiment, and the corresponding code specifications written in four different chart grammars. Here, we present images rendered with \lib and verify that the competing grammars produce nearly identical visual outputs. }
    \label{fig:human_stimuli}
\end{figure*}

We select four chart types---bar, line, scatter, and pie charts---as they are widely regarded as basic and commonly used chart types in the literature \cite{saket19tvcg, battle18chi, lee17tvcg, grammel10tvcg} (R1, R2). Although geographic maps are also common in practice \cite{battle18chi}, we exclude them because novice designers may find it difficult to reproduce their spatial structure on paper with reasonable fidelity within a limited time (R1).

We then design two charts for each chart type: one is \textit{easy} and the other is \textit{hard} to draw. The easy condition assesses whether participants can understand the core construction of each grammar, whereas the hard condition examines how interpretability changes as additional visual encodings and specification components are introduced.
Specifically, each easy example contains only two visual encodings (e.g., $x$ and $y$ axes), while the corresponding hard example introduces one additional encoding or a feature (e.g., regression line).
To keep the chart design grammar-agnostic (R3), we first create each target visualization by hand, without referring to any particular grammar. We then manually prepare the sample data and a natural language prompt describing the charts. Finally, we provide the same prompt and data to an LLM and independently generate specifications in each of the four grammars.
As in our previous experiment (\autoref{sec:machine}), we use GPT-5.6 Sol with ``very high'' reasoning capability to maximize the correctness of the generated specifications. 
We instruct the model to produce concise, non-verbose code containing only the components necessary to construct the target visualization. 
We then manually inspect both the generated code and rendered chart to verify that each specification faithfully reproduces the chart that we intended. Finally, we remove non-semantic styling code, such as grid and axis settings, so the resulting snippets primarily reflect the chart's semantic structure. The stimuli we used are depicted in \autoref{fig:human_stimuli}.

% Full-page portrait stimuli, with the caption below and no rotation.
% \clearpage
% \newgeometry{top=35pt, bottom=10pt, left=10pt, right=20pt}
% \thispagestyle{empty}
% \newsavebox{\humanstimulicapbox}
% \savebox{\humanstimulicapbox}{%
  % Align with the visible chart/code bounds, excluding PDF whitespace.
  % \hspace*{22.3pt}%
%   \begin{minipage}{\dimexpr\linewidth-51.1pt\relax}
%     \captionsetup{font=normalsize,justification=justified,singlelinecheck=false,skip=0pt}
%     \captionof{figure}{The charts that participants are asked to reproduce by hand in our human experiment, and the corresponding code specifications written in four different chart grammars. Here, we present images rendered with \lib, where we also verify that the competing grammars produce nearly identical visual outputs. The rendering results of other grammars are depicted in Appendix C. }
%     \label{fig:human_stimuli}
%   \end{minipage}%
%   % \hspace*{28.8pt}%
% % }
% % \noindent
% \begin{minipage}[c][0.99\textheight][c]{\linewidth}
%   \centering
%   \includegraphics[width=\linewidth,height=\dimexpr0.99\textheight-\ht\humanstimulicapbox-\dp\humanstimulicapbox-8pt\relax,keepaspectratio]{figures/human_stimuli.pdf}\par
%   \vspace{8pt}
%   \usebox{\humanstimulicapbox}
% \end{minipage}
% % \clearpage
% \restoregeometry

\myparagraph{Procedure}
After providing consent, participants receive a brief recap of the study purpose and an explanation of the overall procedure. 
They then complete eight chart drawing trials. Each participant encounters all four grammars, with two trials per grammar---one \textit{easy} and one \textit{hard}.
We counterbalance the order of the four grammars, chart types, and difficulty conditions using separate Latin-square designs.
Participants are given up to 10 minutes to complete each drawing. After completing all eight trials, they participate in a post-study interview in which they mainly describe their perceived comprehensibility of each grammar.
We compensate participants with an equivalent of 20 USD. 
All experiments ended within 60 minutes.

\myparagraph{Study setup}
We design an interface that displays the sample data and corresponding code snippet on the right, with a chat interface on the left through which participants can ask natural-language questions about the functionality of the grammar.
To prevent participants from directly asking how to reproduce a target visualization, we constrain the chat assistant through a system prompt to avoid providing concrete instructions for constructing the visualization (see Appendix C for the full prompt). To minimize interaction latency, we use GPT-5.6-terra with reasoning disabled. 
To improve the accuracy of its responses, the chat interface directly retrieves the documentation of each grammar through the Context7 MCP server.

For drawing, we provide a set of 12 colored pencils, 12 colored markers, and a black ballpoint pen, allowing participants to freely sketch outlines and add color as needed. We guide participants to draw charts on separate A4 sheets and provide unlimited sheets.

% ---------------------

% We print all chart specifications on paper so that participants do not need to alternate their attention between a computer screen and the drawing surface. Each visualization is drawn on a separate A4 sheet, and we provide unlimited sheets. We provide a set of 12 colored pencils and a black pen, allowing participants to freely sketch outlines and add color as needed. While drawing charts, we also allow users to explore each library's official documentation on a separate laptop.

\myparagraph{Expert evaluation}
We recruit four visualization experts (aged 28–51 years; \(M = 35.75\), \(SD = 10.63\); all men). They have between 5 and 12 years of visualization research experience (\(M = 7.50\), \(SD = 3.11\)), hold a Ph.D. in computer science, and regularly publish papers at major visualization venues, such as IEEE TVCG, IEEE VIS, and CGF.
Our study yields 16 (participants) $\times$ 8 (charts per participant) $=128$ drawings in total. We ask each expert to evaluate all drawings.
For each drawing, we present the expert with the corresponding reference chart rendered from the same grammar specification that the participant used to produce the drawing. Experts then assess the quality of each drawing on a Likert scale by answering the following two questions:
\begin{itemize}[leftmargin=9pt]
    \item \textit{Completeness:} Does this drawing include all essential visual elements shown in the reference chart?
    \item \textit{Accuracy:} Does this drawing accurately reproduce the visual representations of the reference chart without redundancy or error?
\end{itemize}
We define these two evaluation criteria based on the recommendations of B\"orner et al. \cite{borner19pnas} for assessing visualization construction.
We instruct experts not to judge paper quality based on stylistic or semantically irrelevant visual characteristics (e.g., color palette, scale, or point size).

\myparagraph{Measurements}
We also record the questions participants ask through the chat interface and the time required to reproduce each chart as additional measures of interpretability. For the questions, we first compare their frequency across grammars as an overall indicator of interpretability (\autoref{sec:quanthuman}). We then qualitatively analyze their content to identify which aspects of each grammar facilitated or hindered participants' understanding (\autoref{sec:humanqual}).

% We recruit four visualization experts (aged XX--YY years; (M = ZZ), (SD = WW); XX men), all of whom hold a Ph.D. and have experience publishing at major visualization venues, such as IEEE TVCG, IEEE VIS, and CGF. Our study yields 16 (participants) $\times$ 8 (charts per participant) $=128$ drawings in total. We ask each expert to evaluate 64 drawings, such that each drawing is independently evaluated by four experts.
% For each drawing, we present the expert with the corresponding reference chart rendered from the same grammar specification that the participant used to produce the drawing. Experts then assess the quality of each drawing on a Likert scale by answering the following two questions:
% \begin{itemize}[leftmargin=9pt]
%     \item \textit{Completeness:} Does this drawing include all essential visual elements shown in the reference chart?
%     \item \textit{Accuracy:} Does this drawing accurately reproduce the visual representations of the reference chart without redundancy or error?
% \end{itemize}
% We define these two evaluation criteria based on the recommendations of B\"orner et al. \cite{borner19pnas} for assessing visualization construction.

\subsection{Quantitative Results}

\label{sec:quanthuman}

We discuss the quantitative findings from our experiment.

\begin{figure*}
    \centering
    \includegraphics[width=\linewidth]{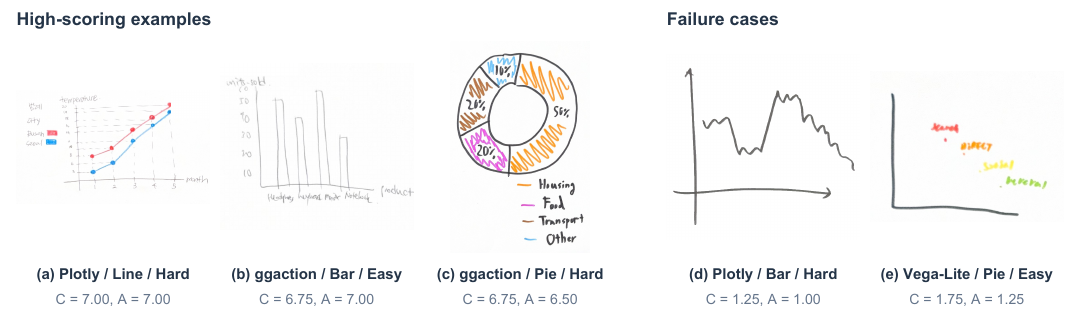}
    \caption{Example drawings that received high (left) and low (right) ratings from expert evaluators. Overall, drawings created from Plotly and \lib{} specifications were rated higher in quality than those created from ggplot2 and Vega-Lite specifications.}
    \label{fig:drawings}
\end{figure*}

\myparagraph{Chart quality} We analyze completeness and accuracy of drawings (examples in \autoref{fig:drawings}) as separate outcomes. Because both measures are rated on a 7-point ordinal scale, we fit a cumulative link mixed model (CLMM) \cite{taylor23brm} separately to each outcome. We model:
\begin{equation*}
\label{eq:clmm}
\begin{aligned}
\mathtt{rating}
&\sim \mathtt{grammar} \times \mathtt{difficulty} \\
&\quad + \mathtt{chartType} + \mathtt{expert} \\
&\quad + (1 \mid \mathtt{participant})
       + (1 \mid \mathtt{drawing}).
\end{aligned}
\end{equation*}
We include an expert term to account for systematic differences in rating severity across expert evaluators. We also include random intercepts for participant and individual drawing to account for repeated drawings produced by the same participant and multiple expert evaluations of the same drawing, respectively.
The inter-rater agreement among experts, measured by the intraclass correlation coefficient, is 0.834.

\begin{figure*}
    \centering
    \includegraphics[width=\linewidth]{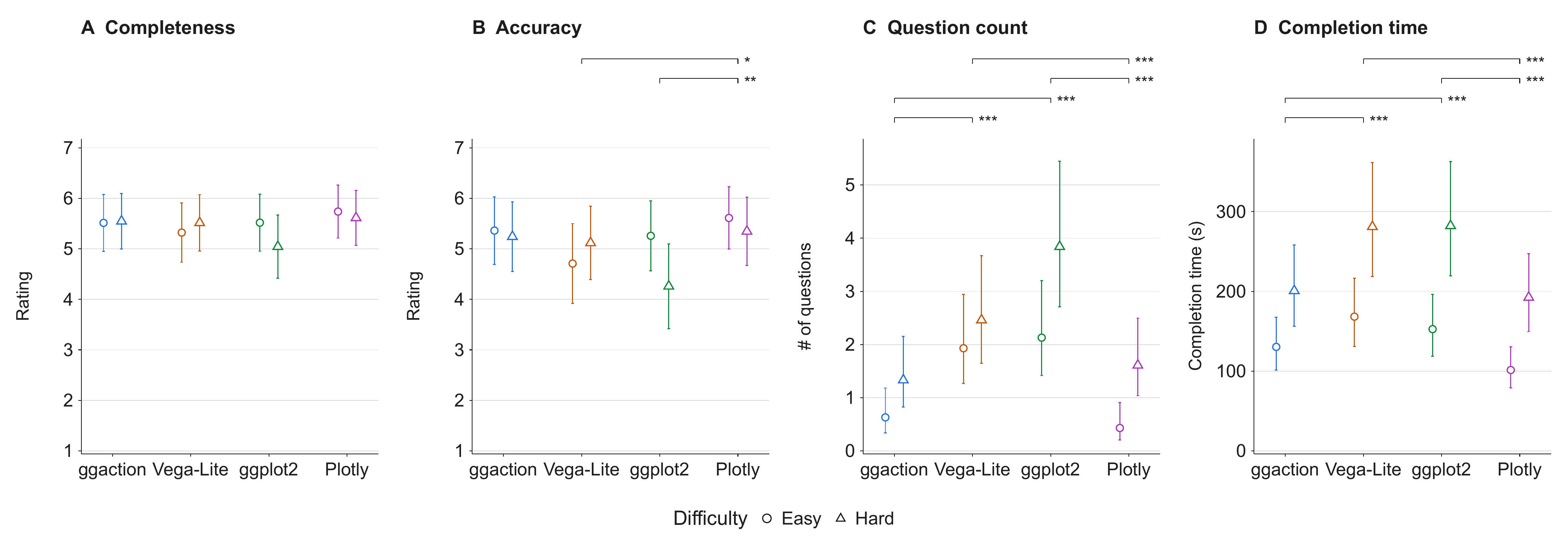}
    \caption{The quantitative results of our user study on the human interpretability of different chart grammars. Although the quality of participants' drawings did not differ that much across grammars (A, B), participants experienced significantly lower interpretive burden with Plotly and \lib (C, D).}
    \label{fig:human_exp}
\end{figure*}

\autoref{fig:human_exp} A and B depict the results.
In terms of completeness, we find no significant effect of grammar ($\chi^2(3)=3.65$, $p=.302$), difficulty ($\chi^2(1)=0.36$, $p=.547$), and their interaction ($\chi^2(3)=2.25$, $p=.523$).
For accuracy, we find a significant effect of grammar ($\chi^2(3)=9.23$, $p < .05$), but neither for difficulty ($\chi^2(1)=1.67$, $p=.196$) nor for their interaction ($\chi^2(3)=5.61$, $p=.132$).
The post-hoc analysis comparing cumulative odds shows that Plotly is likely to have higher accuracy than ggplot2 ($\mathrm{OR}=2.81$, $p < .01$) and Vega-Lite ($\mathrm{OR}=2.34$, $p < .05$), but no other significant differences are observed.
Note that the comparison between \lib and ggplot2
approached, but did not reach, the conventional significance threshold ($\mathrm{OR}=2.11$, $p=.052$).

\myparagraph{Question count}
We analyze how the number of questions asked by participants varies with visualization grammar and task difficulty. Because the dependent variable is a nonnegative count variable that includes zero-valued observations, we use a Poisson generalized linear mixed model (GLMM) with a log link, which ensures that the expected question count remains nonnegative. We fit the model:
\begin{equation*}
\label{eq:glmm}
\begin{aligned}
\log(\mathtt{count})
&\sim \mathtt{grammar} \times \mathtt{difficulty} \\
&\quad + \mathtt{chartType} + \mathtt{trialOrder} \\
&\quad + (1 \mid \mathtt{participant}).
\end{aligned}
\end{equation*}
We include trial order to account for potential ordering effects. A random intercept for participant accounts for repeated trials completed by the same participant.

As a result (\autoref{fig:human_exp} C), the number of questions differs significantly across both grammars ($\chi^2(3)=53.05$, $p<.001$) and difficulty ($\chi^2(1)=22.93$, $p<.001$), where there is no interaction effect ($\chi^2(3)=5.01$, $p=.171$).
Post-hoc pairwise comparisons
reveal that both \lib and Plotly elicit fewer questions than ggplot2
and Vega-Lite ($p<.001$ for all). We find no significant difference between
\lib{} and Plotly ($p=.721$).

\myparagraph{Completion time}
We analyze how task completion time is affected by grammar and task difficulty. 
As completion times are positive and right-skewed, we analyze
log-transformed completion time using a linear mixed-effects model:
\begin{equation*}
\begin{aligned}
&\log(\mathtt{completionTime}) \\
&\quad \sim \mathtt{grammar} \times \mathtt{difficulty} \\
&\qquad + \mathtt{chartType} + \mathtt{trialOrder} \\
&\qquad + (1 \mid \mathtt{participant}).
\end{aligned}
\end{equation*}
Note that we use the same model variables with the question count, except for the dependent variable.

As a result (\autoref{fig:human_exp} D), we find significant main effects of both grammar
($\chi^2(3)=48.49$, $p<.001$) and difficulty
($\chi^2(1)=111.38$, $p<.001$) on completion time, where hard trials take longer to complete.
We find no significant interaction between grammar
and difficulty ($\chi^2(3)=2.26$, $p=.521$).
Post-hoc comparisons show that participants complete trials with \lib{} faster than with ggplot2 ($\mathrm{ratio}=0.779$, $p<.001$), and
than with Vega-Lite ($\mathrm{ratio}=0.744$, $p<.001$).
Plotly is also faster than both ggplot2 and Vega-Lite
(both $p<.001$), whereas we find no other significant differences between grammars.

\subsection{Qualitative findings}

\label{sec:humanqual}

We qualitatively examine participants' questions to the chat assistant and their post-study interviews. 

\myparagraph{Familiar names helped participants}
Participants used familiar chart terms as cues when interpreting specifications. P4 explained that the \codeverb{createBarPlot} and \codeverb{createScatterPlot} operations in \lib{} make the intended actions recognizable, and described a similar
benefit from Plotly's \codeverb{pie} and \codeverb{line} functions. Unfamiliar terminology, by contrast, introduced additional decoding effort. 
P13 reported frustration when facing ggplot2's \codeverb{aes} and \codeverb{geom}
functions. Consistently,
13 of the 16 participants explicitly asked about these two in their chat messages.
Similarly, Vega-Lite's data type declarations (e.g., \codevalue{quantitative} and \codevalue{nominal}) also raised uncertainty.

\myparagraph{Encodings beyond the $x$ and $y$ axes are difficult to interpret}
Participants described \texttt{x} and \texttt{y} texts as useful anchors for connecting data fields to axes, but remained uncertain about less familiar properties. For example, P13 reported that she immediately recognized the chart type and axis mappings in \lib{}'s \codeverb{createBarPlot} action, yet hesitated over whether its \codevalue{group} layout implied grouped or stacked bars.
Similarly, P8 understood the color grouping in a Vega-Lite specification but was unsure whether \codeparam{xOffset} required adjacent bars or separate displays.

\myparagraph{The verb--noun schema for naming \lib functions improves interpretability}
Participants generally found \lib's function names easy to understand. For example, P8 appreciated that \lib explicitly indicates the action to be performed, suggesting that verbs in their names helped participants infer the intended operation from the code. However, recognizing an operation did not necessarily clarify all its details, particularly its parameters: P8 remained uncertain about the meaning of \codeparam{band}\texttt{: }\codevalue{false}, while P16 wanted additional information about the regression model in use. By contrast, some participants found Plotly's abstractions overly opaque, noting that they occasionally had to pause to determine what the code was doing.

\subsection{Discussions}

\myparagraph{\lib{} lowers the effort required to interpret chart specifications}
Participants asked fewer questions and completed tasks faster with \lib{} than with ggplot2 and Vega-Lite, while its performance was comparable to Plotly. Qualitative feedback further suggests that \lib{} provides abstractions that more closely align with how participants reason about chart authoring than Plotly. 
However, we find no statistically significant advantage in completeness or accuracy. Thus, our results do not show that \lib{} enables users to understand charts more accurately; rather, it reduces the effort required to reach that understanding. Participants were able to recover the intended charts given sufficient time, regardless of grammar; \lib{} shortened this process by reducing the conceptual distance between familiar chart concepts and code specifications.

\myparagraph{\lib{} does not always expose the detailed visual consequences}
The verb--noun naming scheme of \lib{} helped participants recognize what an operation was intended to do, but did not necessarily reveal exactly how that operation would affect the resulting chart. 
For example, \codeverb{createRegression} clearly indicates that a regression line is added, yet participants still questioned whether an uncertainty band was included or not. Thus, \lib{} does not make every downstream visual consequence of functions or parameters equally apparent. 
This suggests that consequential design choices may need to be surfaced more explicitly in the specification, a tradeoff we revisit in \autoref{sec:framing}.
\section{Discussions}

Our research raises broader questions about how \lib and future grammars should be designed and used.

\subsection{How Much Should the Grammar Express?}

\label{sec:tradeoff}

Our expressiveness evaluation (\autoref{sec:expresiveness}) shows that \lib{} covers most chart authoring intents observed in practice, while some remain only partially supported or unsupported. This raises a broader question: how should \lib{} close these remaining gaps?
Adding actions can increase expressiveness and make it easier for designers or language models to map a given intent to an appropriate operation. 
However, extending the action vocabulary also introduces costs. 
A larger vocabulary increases the number of functions that must be distinguished, particularly when multiple actions have similar semantics. It also increases implementation complexity: as discussed in \autoref{sec:techdetail}, implementing an action in \lib{} already requires not only modifying the underlying state but also managing the rematerialization of dependent graphics. Thus, adding actions does not simply eliminate complexity. Rather, it can shift complexity from repeatedly composing low-level operations at use time into the grammar and its implementation.

From this perspective, a useful criterion for introducing a new action is whether it encapsulates a composition that many users would otherwise have to perform repeatedly. Determining which compositions meet this criterion ultimately requires empirical observation. Just as the initial design of \lib was informed by prior observations of chart authoring processes in the wild \cite{wang23tvcg, srinivasan21chi}, future extensions should examine which authoring intents recur in practice and how users naturally decompose them into actions. Building chart authoring systems on top of \lib and studying their use in realistic settings therefore provides a plausible future direction.

%% expressiveness가 늘어나면 좋지만, complexity가 늘어난다...
%% 어느 액션까지 넣어야하는가. 우리 expresiveness 실험을 보면...

\subsection{Which Part of Intents Deserves to be Explicit?}

\label{sec:framing}

\lib{} currently represents actions through a verb--noun function paired with parameters: the function specifies the primary operation, whereas parameters capture more specific design decisions. In this context, once an intent is deemed worthy of inclusion in the action vocabulary, another question arises: which design decisions warrant a function of their own, and which can remain implicit or parameterized?
Our human subject study (\autoref{sec:humaninter}) illustrates this tension. Participants generally understood \codeverb{createRegression} and were able to draw the intended linear regression, yet some questioned which regression model the function actually applied. This suggests that the implicit default of linear regression, while unproblematic for many users, can obscure a design decision for others. 
When such a decision substantially affects the resulting chart, it may deserve greater visibility in the specification---for example, through an explicit function naming (e.g., \codeverb{createLinearRegression}) rather than an implicit default.

Making more decisions explicit, however, introduces another form of complexity. Beyond increasing the size of the action vocabulary and its associated learning and maintenance costs (\autoref{sec:tradeoff}), excessive explicitness can make individual specifications more verbose. This can increase token usage for language models and the amount of information human designers need to inspect. 
The challenge is therefore not to make every design decision explicit, but to determine which ones are important enough to warrant explicit attention.

Answering this question again requires empirical observation, but of a different kind from that discussed above (\autoref{sec:tradeoff}). Rather than identifying which authoring intents recur in practice, we need to understand which design decisions users treat as natural defaults and which they expect to be stated explicitly. Observing where users hesitate, ask for clarification, or misinterpret defaults in realistic authoring workflows may provide an empirical basis for deciding how \lib should distribute design intents among functions, parameters, and default settings.

% 중요한 선택은 코드에서 좀 더 눈에 띄게 보여줄 필요가 있다.

% 예를 들어 band=false가 실제 차트 결과에 중요한 영향을 준다면, 그냥 작은 parameter로 숨기기보다 더 이해하기 쉬운 이름을 쓰거나 아예 별도 함수로 만드는 것도 고려할 수 있다는 거지.

% createRegression(...)
% hideRegressionBand(...)

% 이런 식으로.

% 물론 모든 parameter를 함수로 만들면 코드가 너무 길어지니까, 핵심 질문은:

% 어떤 세부사항까지 함수 이름으로 명확하게 보여주고, 어떤 것은 parameter로 남길 것인가?

\subsection{Who Will Visualization Grammars Be For?}

Our experiments show that \lib{} is easier for SLMs to generate than competing grammars (\autoref{sec:machine}). Today, the value of such machine-friendliness is straightforward: a representation that can be generated reliably by smaller models reduces the computational cost. Human interpretability is similarly important because, as our machine experiment also demonstrates, smaller models still make errors. Readable specifications therefore allow users to inspect and correct machine-generated outputs.

This motivation, however, may weaken as language models improve. In the near future, models may generate Vega-Lite, ggplot2, \lib{}, or other grammars nearly perfectly, quickly, and at negligible cost. 
If so, does a visualization grammar still need to be machine-friendly? Likewise, if models can generate perfect code, does it still need to be human-readable?

This leads to an ultimate question: \textit{who is a visualization grammar ultimately for?} One dystopian future is that visualization code becomes entirely invisible. Users interact with agents only through natural language and rendered charts, while agents internally produce whatever representation is convenient. 
In this scenario, grammars need not be easy for humans to read, closely aligned with natural-language intent, or maybe even efficient for models to generate. 
In such a future, many advantages of \lib that we verified may indeed become less important. 

A second possibility, however, provides \lib a different and potentially longer-lasting role. Perfectly generating code is not the same as perfectly understanding a user's design intent. Visualization authoring involves many decisions for which there is no single correct answer. 
Users may not fully know or articulate their intentions in advance, and those intentions often evolve during the authoring process (\autoref{sec:chartauthoringprocess}). 
Thus, even if an agent can generate syntactically valid programs without errors, mismatches between the agent's design decisions and the user's evolving intent need not disappear. 
From this perspective, we argue that the role of \lib{} may need to be reframed in the near future. Rather than viewing it merely as a grammar that is easy for machines to generate and humans to read, we may instead view it as a representation of design decisions that facilitates collaboration between humans and agents in chart authoring.

% In this view, the lasting goal \lib may not lie in making code easier for humans and machines to interpret, but in facilitating their collaboration.

% keeping design decisions visible as control passes between them. The question, then, is not simply whether visualization grammars should be designed for humans or machines, but whether they can serve as a shared medium through which people inspect, challenge, and revise the decisions made by agents. Whether \lib{} actually improves such human--agent collaboration remains an important question for future work.

% % inspection, discussion, revision, and negotiation. Visualization code then shifts from being primarily a \emph{programming artifact} to also serving as a \emph{decision artifact} in human--agent collaboration.

% This raises a broader question for the future of visualization grammars: \textit{will visualization code disappear behind increasingly capable agents, or will it remain visible as an interface through which humans and agents negotiate design intent?} If the latter, the long-term value of representations such as \lib{} may lie less in making code generation easier and more in making the agent's design reasoning legible and revisable.

\section{Conclusion}

In this paper, we propose \lib, a visualization grammar that represents charts as a sequence of authoring actions.
By doing so, \lib makes code specifications more aligned with the chart authoring intents of designers. 
By conducting a series of evaluations, we find that \lib is sufficiently expressive and interpretable to both humans and machines. 

At its core, \lib{} explores a simple shift in abstraction: treating design intent, rather than graphical structure, as a first-class unit of visualization specification. Our results suggest that this shift is useful for both interpreting and generating chart specifications. We hope this perspective opens new opportunities for designing visualization grammars and authoring systems around the decisions people actually make in real-world authoring processes. 

\bibliographystyle{ACM-Reference-Format}
\bibliography{ref}

% \appendix









\end{document}